\documentclass[structabstract]{aa}  
\usepackage{graphicx}
\usepackage{txfonts}

\title{H$\alpha3$: an H$\alpha$  imaging survey of HI selected galaxies from ALFALFA
\thanks{Observations taken at the  observatory of San Pedro Martir (Baja California, Mexico), belonging to the
Mexican Observatorio Astron\'omico Nacional.}}
\subtitle{IV. Structure of galaxies in the Local and Coma Superclusters
\thanks{Tables A1 and A2 are only available in electronic form at the CDS via anonymous ftp to 
cdsarc.u-strasbg.fr (130.79.128.5) or via http://cdsweb.u-strasbg.fr/cgi-bin/qcat?J/A+A/ }}

\author{M. Fossati \inst{1,2}
\and G. Gavazzi \inst{2}
\and G. Savorgnan \inst{3}
\and M. Fumagalli \inst{4,5} \thanks{Hubble Fellow}
\and A. Boselli \inst{6}
\and L. Guti{\'e}rrez \inst{7}
\and H. Hern{\'a}ndez Toledo \inst{8}
\and R. Giovanelli \inst{9}
\and M.P. Haynes \inst{9}
}

\authorrunning{M. Fossati et al.}
\titlerunning{H$\alpha3$: H$\alpha$ imaging survey of HI selected galaxies from ALFALFA}

\institute{Max-Planck-Institut f{\"u}r Extraterrestrische Physik, Giessenbachstrasse, D-85748 Garching, Germany\\
\email {mfossati@mpe.mpg.de}
\and 
Universit\`a degli Studi di Milano-Bicocca, Piazza della Scienza 3, 20126 Milano, Italy\\
\email {giuseppe.gavazzi@mib.infn.it}
\and
Centre for Astrophysics and Supercomputing, Swinburne University of Technology, Hawthorn, Victoria 3122, Australia\\
\email {gsavorgn@astro.swin.edu.au}
\and
Carnegie Observatories, 813 Santa Barbara Street, Pasadena, CA 91101, USA\\
\email {mfumagalli@obs.carnegiescience.edu}
\and
Department of Astrophysics, Princeton University, Princeton, NJ 08544-1001, USA
\and
Aix Marseille Universit{\'e}, CNRS, LAM (Laboratoire d'Astrophysique de Marseille) UMR 7326, 13388, Marseille, France\\
\email {alessandro.boselli@oamp.fr}
\and 
Instituto de Astronom\'ia, Universidad Nacional A\'utonoma de M\'exico, Carretera Tijuana-Ensenada, km 103, 22860 Ensenada, \\
B.C. M\'exico \email {leonel@astro.unam.mx}
\and
Instituto de Astronom\'ia, Universidad Nacional A\'utonoma de M\'exico, Apartado Postal 70-264, 04510 M\'exico D.F., M\'exico\\
\email {hector@astroscu.unam.mx}
\and
Center for Radiophysics and Space Research, Space Science Building, Ithaca, NY 14853, USA\\
\email {haynes@astro.cornell.edu, riccardo@astro.cornell.edu}
}

\begin{document}
   
\date{Received 14 December 2012 / Accepted 3 March 2013}

 \abstract
%Context
{We present the analysis of the galaxy structural parameters from H$\alpha3$, 
 an H$\alpha$ narrow-band imaging follow-up survey of $\sim 800$ 
 galaxies selected from the HI  Arecibo Legacy Fast ALFA Survey (ALFALFA) 
 in the Local Supercluster, including the Virgo cluster, and in the
 Coma Supercluster.}
{Taking advantage of H$\alpha3$, which provides the complete census 
 of the recent star-forming, HI-rich galaxies in the local universe,
 we aim to investigate the structural parameters of the young ($<10$ Myr) and
 the old ($>1$ Gyr) stellar
 populations. By comparing the sizes of these stellar components,
 we investigated the spatial scale on which galaxies are growing at the present cosmological epoch
 and the role of the environment in quenching the star-formation activity.}
{We  computed the concentration, asymmetry, and clumpiness (CAS) structural parameters for recently born and old stars.
To quantify the sizes we computed half-light radii and a new parameter dubbed $EW/r$ based on the half-light radius of the 
H$\alpha$ equivalent width map.
To highlight the environmental perturbation, we adopt an updated calibration of 
the HI-deficiency parameter ($Def_{HI}$)  that we use to divide the sample in 
unperturbed galaxies ($Def_{HI} \leq 0.3$) and perturbed galaxies ($Def_{HI}>0.3$).}
{The concentration index computed in the $r$ band depends on the stellar mass and on the Hubble type these variables 
are related because most massive galaxies are bulge dominated therefore highly concentrated. 
Going toward later spirals and irregulars the concentration index and the mass decrease along with the bulge-to-disk 
ratio. Blue compact dwarfs (BCDs) are an exception because they have similar mass, but they are more concentrated 
than dwarf irregulars. The asymmetry and the clumpiness increase along the spiral sequence up to Sc-Sd,
but they decrease going in the dwarf regime, where the light distribution is smooth and more symmetric.
When measured on $\rm{H\alpha}$ images, the CAS parameters show no obvious correlations with Hubble type.
Irrespective of whether we used the ratio between effective radii or the $EW/r$ parameter, 
we found that the concentration index is the main parameter that describes the current growth of isolated galaxies but, for a fixed 
concentration, the stellar mass plays a second-order role. At the present epoch, massive galaxies are growing
inside-out, conversely, the dwarfs are growing on the scale of their already assembled mass.}
	 {}    


\maketitle

\section{Introduction}
Understanding the physical structure of galaxies is a key to shed light on the assembly of the baryonic matter in the Universe during 
the cosmic growth of large-scale structures. The first and most widely used way to describe the physical structure of galaxies
is the morphological classification. Since the 1920s several morphological schemes have been proposed 
and adopted (Hubble 1936; de Vaucouleurs 1959). Although the adopted criteria were somewhat subjective, noticeably
the different classifications proposed by the most distinguished astronomers were consistent.
In recent years, the enormous size of galaxy samples (e.g., those produced by the Sloan Digital Sky
Survey, SDSS) made the visual estimate of the morphology  
virtually impossible (with the exception of the GalaxyZoo project by Lintott et al. (2008), which aims at a human classification using a 
quarter of million of volunteers). Nevertheless, the improvements of computer science made possible the advent of automated 
quantitative morphological classification methods, which are trying to make the scheme proposed by Hubble ``quantitative''
using objective parameters (e.g. Cheng et al. 2011).
One alternative is to describe a galaxy parametrically, by modeling the distribution of light as 
projected onto the plane of the sky with a prescribed suite of analytic functions. 
Following this parametric modeling, the light profile can be decomposed into a bulge and a disk that can be fitted with a Sersic 
function (Sersic 1968) and an exponential, respectively. However, Laurikainen et al. (2005) and Peng et al. (2010) have shown that 
simple bulge + disk models are inadequate for dwarf and irregular galaxies. 

More recently, a new non-parametric approach to galaxy morphological classification is emerging.
This scheme does not assume a particular analytic function for the 
galaxy's light distribution and therefore may  apply to irregulars as well as to galaxies on the
Hubble sequence (Lotz et al. 2004).
In addition, these methods are less computationally intensive than profile fitting and can be used even at high redshift
where the resolution and signal-to-noise are too low for a parametric (or even visual) classification. 
One of the most popular schemes is the concentration, asymmetry, and clumpiness (CAS) system developed by Conselice (2003), inspired 
by other seminal works (Abraham et al. 1996; Takamiya 1999; Bershady et al. 2000; Conselice et al 2000).
Although several works use the CAS classification at high redshift (e.g. Conselice et al. 2003, 2007, 2008; Cassata et al. 2005), systematic attempts to quantify the CAS parameters at low redshift are 
rare in the literature. Previous efforts have been carried out by Hern{\'a}ndez-Toledo et al. (2008), applied to isolated galaxies
selected from the CIG catalog (Karachentseva 1973). The first goal of the present work is to produce a reference for the CAS parameters
computed on a well-selected, complete sample that is representative of late-type galaxies at $z\sim0$.

Since Walter Baade (1944) discovered that young Pop-I stars have a broader spatial distribution than older Pop-II in spiral nebulae 
(made possible by wartime blackouts, which reduced the light pollution at Mount Wilson),
it became increasingly clear that galaxies grow inside-out (see e.g. Mu{\~n}oz-Mateos et al. 2007).
Owing to the extensive sky coverage of the SDSS survey (Abazajian et al. 2009) and of the GALEX (Martin et al. 2005) mission,
even an eyeball inspection of pseudo-color images of galaxies, especially the most inclined ones in the plane of the sky,
reveal bluer colors in the outer wings, suggesting younger stellar populations (see e.g. de Jong 1996).
However, a caveat is that radial gradients of metallicity (Skillman et al. 1996) and dust extinction would eventually artificially enhance  
any color gradient induced by a pure change in stellar age. Resolved stellar populations have been studied with the Hubble space 
telescope (HST) and revealed, for example, the inside-out growth of the disk galaxy M33 (Williams et al. 2009).
A definitive proof of inside-out build-up of galaxies would need to be derived from a direct comparison
of the radial extent of old stars (traced for example by $r$-band observations or redder)
with that of stars as young as $10^{6-7}$ yr, such as those revealed by H$\alpha$ observations.
Ground-based integral-field spectroscopy has made this possible and demonstrated the power of spatially resolved
emission line surveys at low- and high-redshift (e.g. F{\"o}rster Schreiber et al. 2009; Law et al. 2009; Sanchez et al. 2012), 
but it has been limited to small samples due to the required extensive observational effort. 
Recently, a comparison between the $r$- and H$\alpha$ scale-lengths of galaxies 
at $z=1$ was carried out by Nelson et al. (2012), using spatially resolved HST grism spectra from the 
3D-HST survey (Brammer et al. 2012), showing that the effect is mild at best. In addition to the CAS parameters, we present here 
the size measurements of local star-forming galaxies as traced by different stellar populations.

The outline of the paper is as follows: Section \ref{samplessect} introduces the selection criteria and the completeness of the samples
used in this work. In Section \ref{CASsect}, the CAS parameters are reviewed, together with a description of computation details and 
validation tests for the specific software we developed for this paper. In Section \ref{sizes}  the methods used for the size 
measurements of the spatial extent of galaxies are presented. Finally, the discussion of our results follows in Section \ref{resultsect}.
Throughout the paper we adopt a flat $\Lambda CDM$ cosmology, with  
$H_0=73 \rm ~km~s^{-1}~Mpc^{-1}$ and $\Omega_\Lambda=0.7$.
All magnitudes are given in the AB system unless explicitly noted otherwise.

\section{Samples} \label{samplessect}
\subsection{Local Supercluster} \label{Virgo}
The first sample analyzed in this work is drawn from the 900-square-degrees region between
$\rm 11^h <R.A. <16^h$ and $4^o< Dec. <16^o$, which covers the Local Supercluster, including the Virgo cluster. 
This region has been fully mapped by ALFALFA (Giovanelli et al. 2005), which provides us with a complete sample 
of HI-selected galaxies with masses as low as $10^{7.7} ~\rm M_\odot$ (Haynes et al. 2011).

\begin{figure*} [!t]
 \centering
\includegraphics[scale = 0.85]{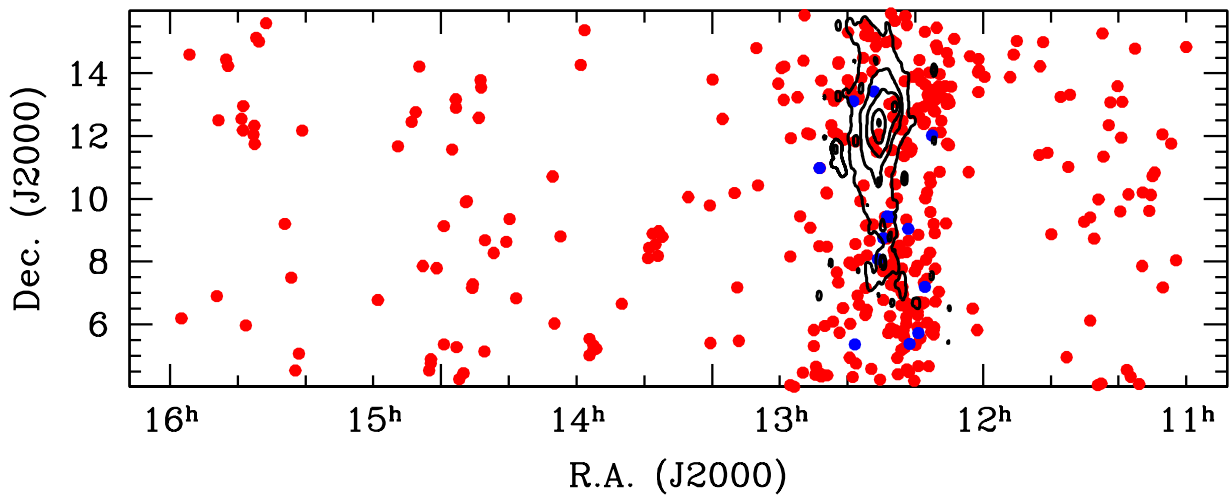}
\caption{Sky distribution (R.A. and Dec. not in scale) of the Local Supercluster galaxies with superposed
the X-ray contours (B{\"o}hringer et al. 1994) highlighting the position of the Virgo cluster.
Red points refer to the 372 late-type galaxies observed by $\rm{H\alpha}3$ in the ALFALFA 
footprint. The 13 optically selected VCC galaxies followed-up in $\rm{H\alpha}$ are shown as blue points.}
\label{skyLocal}
\end{figure*}

H$\alpha3$ is a follow-up survey consisting of 
H$\alpha$-imaging observations of ALFALFA detections with high signal-to-noise (typically $S/N>6.5$)
and a good match of two independent polarizations (code = 1 sources; Giovanelli et al. 2005, 
Haynes et al. 2011). H$\alpha$ observations of members of the Virgo cluster were presented in  
Gavazzi et al. (2002a,b, 2006), Boselli \& Gavazzi  (2002), and Boselli et al. (2002a,b). 
Images and fluxes are also publicly available via the GOLDMine web server (Gavazzi et al. 2003).
H$\alpha$ data for galaxies in low-density regions of the Local Supercluster were presented in 
Gavazzi et al. (2012 hereafter Paper I).
Outside the Virgo cluster the redshift window\footnote{The 
lower velocity limit is set to avoid galaxies whose H$\alpha$ line falls 
on the steep shoulder of the transmission curve of the filter used at SPM (see Paper I).}
 of H$\alpha3$ is $ 350<cz < 2000 \rm ~km~s^{-1}$, and
we limit our selection to objects with HI fluxes $F_{\rm HI} > 0.7 {\rm~Jy~km~s^{-1}}$,
while in the Virgo cluster the velocity interval 
is extended to $ 350<cz < 3000 \rm ~km~s^{-1}$, without any constraint on the HI flux, 
to map the cluster in its full extent.
At the distance of 17 Mpc assumed for the Virgo subcluster A (Gavazzi et al.
1999), the adopted flux limit corresponds to an HI mass $M_{\rm HI}=10^{7.7} ~M_\odot$, 
while H$\alpha3$ is complete to $M_{HI}>10^{8}~ M_\odot$ (see Paper I). 
ALFALFA detects about $50\%$ of the late-type galaxies in the Virgo cluster, either because the 
remaining are dwarf galaxies with an HI content below the limiting sensitivity or because they are
highly HI-deficient galaxies. 
Fig. \ref{skyLocal} illustrates the sky region covered by H$\alpha3$, which contains 409 galaxies. Among these, 372 objects are
late-type galaxies and are shown as red dots. 

In addition, 13 galaxies from the Virgo Cluster Catalog (VCC, Binggeli et al. 1985) that are undetected by ALFALFA have been 
previously observed in H$\alpha$. They are well-known highly HI-deficient objects from previous deep pointed observations 
taken at Arecibo by Helou et al. (1984), Haynes \& Giovanelli (1986), Hoffman et al. (1987, 1989), and Gavazzi et al. (2005).
Although these galaxies are not selected according to any criterion of completeness, we included them to fully exploit 
the dependence of the structural parameters on the amount of HI. These galaxies are shown in Fig. \ref{skyLocal} as blue dots.
The total number of late-type galaxies in this sample is 385.

\subsection{Coma Supercluster}
This second sample includes $\sim 650$ nearby galaxies, selected from ALFALFA in the ``spring'' sky 
of the Coma Supercluster. The traditional area of the Coma Supercluster ($\rm 11.5^h < R.A. <13.5^h\, ; \,18^o< Dec. <32^o\, 
; \,3900<cz<9500$, hereafter region 1) is a very complex environment composed of two major 
clusters, Coma and A1367, and smaller groups and filaments that connect the clusters and build 
the Great Wall (Gavazzi et al. 2010). Only a small strip (hereafter region 2) between $\rm 10^h <R.A. <16^h$,  
$24^\circ< Dec. <28^\circ$, and $3900<cz<9500\rm{~km~s^{-1}}$ is currently included in the
ALFALFA catalog (Haynes et al. 2011). At present, half of the Coma cluster lies in the footprint of ALFALFA. 
We found 683 galaxies with high S/N and a good match of two independent polarizations. 
These are the same criteria as for the Local Supercluster selection but, because of the 
distance of Coma (100Mpc) we applied no additional threshold on the limiting HI flux.   
The limiting sensitivity of ALFALFA at the distance of Coma is $ M_{\rm HI~lim} = 10^{9.1} M_\odot$. 
Furthermore, the $\rm{H\alpha}3$ survey is still not complete in this strip, where only 
261 galaxies have been followed-up in $\rm{H\alpha}$ at present. 

Because neither ALFALFA nor $\rm{H\alpha}3$ are currently complete in the whole region under study,
we complemented the HI selection with galaxies optically selected from the 
CGCG catalog (Zwicky et al. 1961-68) that were previously observed in $\rm{H\alpha}$.
We found 372 bright ($m_{\rm lim} = 15.7~mag_{\rm vega}$) late-type galaxies in the CGCG in region 1.
Of these, 82 are in the region covered by ALFALFA and are HI-detected therefore we refer to them as radio-selected galaxies. 
The remaining 290 galaxies lie beyond the ALFALFA footprint or are HI-undetected by ALFALFA, therefore they are considered to be 
optically selected. The majority of these optically selected galaxies (230) are detected in HI from previously deep pointed
observations (see references in Sec \ref{Virgo}). We recall that because the HI observations have been carried out with different integration 
times, this sample is not complete to a well defined HI-mass limit. Nonetheless, the optically selected galaxies have HI properties that 
mimick those of the radio-detected ones at the distance of Coma, and there is every expectation that most of them ($\sim 80\%$) 
will be detected by ALFALFA. The remaining ($\sim 20\%$) are HI-deficient galaxies, for which ALFALFA is too shallow. 
Our combined sample is at least as deep as ALFALFA for the HI content of the galaxies. We refer to the detailed comparison of the optical 
vs. radio selection performed by Gavazzi et al. (2013b, hereafter Paper III) for the other properties. According to these authors, 
our selection of bright objects from the CGCG catalog allowed us to build a sample whose stellar masses, colors and star formation 
rates are not significantly different from the late-type galaxies selected from ALFALFA. 
In conclusion, we expect to introduce no biases by combining the two selection criteria.  

\begin{figure*} [t]
\centering
\includegraphics[scale = 0.85]{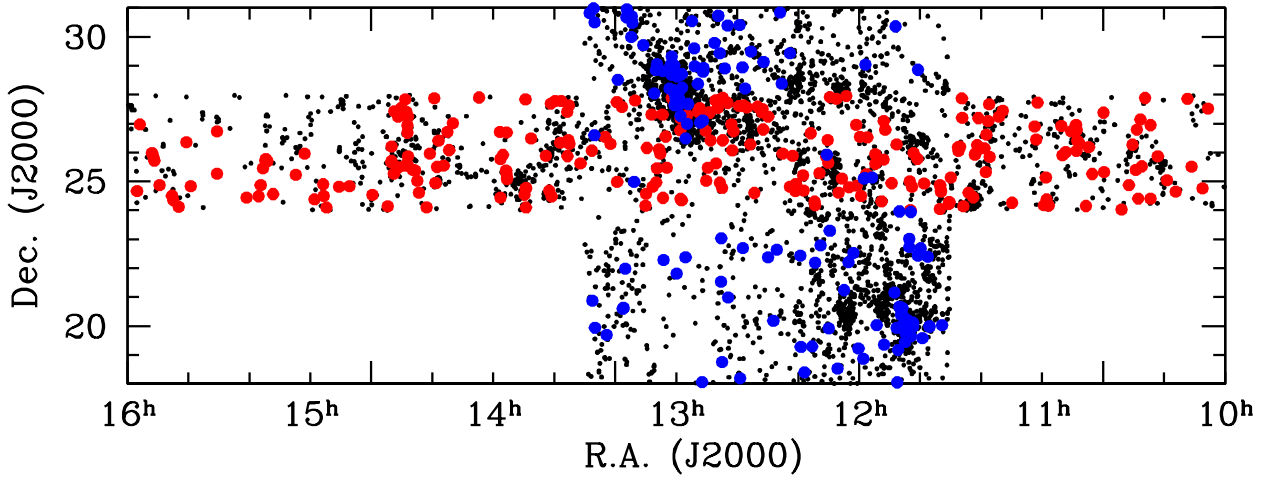}
\caption{Sky distribution (R.A. and Dec. not in scale) of the Coma Supercluster galaxies. 
Black dots represent the optically selected galaxies (all types) selected from the SDSS 
database ($11^h30^m < R.A. < 13^h30^m$, region 1) or from ALFALFA in the strip between $24^\circ < Dec. <28^\circ$
beyond the previous R.A. window (region 2). Red dots refer to the 261 galaxies observed in $\rm{H\alpha}3$ in the ALFALFA 
strip. The 139 optically selected CGCG galaxies followed-up in $\rm{H\alpha}$ outside the ALFALFA strip and 16 
in the ALFALFA strip are shown as blue dots. }
\label{skyComa}
\end{figure*}

The $\rm{H\alpha}$ images in the Coma sample are taken from $\rm{H\alpha}3$ (Gavazzi et al. in prep., hereafter Paper V) 
(see Fig. \ref{skyComa} red points) or from previous observations of 155 optically selected CGCG galaxies by Gavazzi et al. 
(1999, 2002a, 2002b, 2006), Boselli \& Gavazzi (2002), Iglesias-P{\'a}ramo et al. (2002), and Cortese et al. (2006)
(see Fig. \ref{skyComa} blue points).
For the CGCG, images and fluxes are publicly available via the GOLDMine web server (Gavazzi et al. 2003).
The number of galaxies (with available $\rm{H\alpha}$ images) in the Coma Supercluster sample is 413, which brings the total 
number of galaxies analyzed in this work to 798.

Following the prescriptions adopted in the previous papers of the series, the stellar mass was derived from the SDSS $i$ magnitudes and $g-i$ color using 
the transformation 
\begin{equation}
\log \Bigl(\frac{M_{*}}{M_{\odot}} \Bigr) = -1.94 + 0.59 \cdot (g-i) + 1.15 \cdot \log \Bigl(\frac{L_{i}}{L_{\odot}} \Bigr), 
\label{eq:our_mass}
\end{equation} 
where $\log L_{i}$ is the $i$-band luminosity in solar units ($\log L_{i}=(I_{mag}-4.56)/-2.5$). This is a modification of the 
Bell et al. (2003) formula introduced for consistency with the mass determination of MPA-JHU\footnote{http://www.mpa-garching.mpg.de/SDSS/DR7/}.
Morphological types are taken from the VCC and CGCG catalogs when available. For all the others
the type was assigned after visual inspection of the SDSS pseudo-color images. Following a procedure similar to the one described in
Gavazzi et al. (2010), giant late-type galaxies (Sa - Sd) were classified in steps of half Hubble class on
a purely morphological basis. Among dwarf galaxies, blue compact objects (BCD) were classified according to their amorphous 
compact morphology and blue color, on the other hand, dIrr are blue but low surface brightness objects.

\section{The CAS System} \label{CASsect}
 
One of the most popular non-parametric sets of morphological measurements is the CAS system introduced 
by Conselice (2003).
It is based on three indices: concentration  $C$, asymmetry  $A$, and  clumpiness  $S$. 

\subsection{Computing CAS parameters} \label{computingCAS}
Before structural parameters measurements were obtained, basic digital image-processing was applied to all frames
(i.e., bias-subtraction, flat-fielding, cosmic-ray removal, and sky-subtraction; see Paper I). 
In addition, any objects in the galaxy's image that are not part of the galaxy, such as foreground stars 
or foreground and background galaxies, were removed using the {\sc IRAF} task \emph{imedit}.
CAS parameters in the Gunn $r$-band images were computed as described in 
the following subsections for all galaxies except for close pairs or those that are affected by haloes of bright stars.
Furthermore, we disregarded galaxies whose surface brightness is so low that the centering and noise-correction 
routines never converge. These two problems prevented us from measuring structural parameters for 56 galaxies in our
samples.

The CAS parameters were also computed in the $\rm{H\alpha}$ images. The center of the images was set to be the same computed in 
the $r$ band since $\rm{H\alpha}$ is often centrally depressed (Shapley 2011). In addition, the 
concentration index is affected if there is an active galactic nucleus (AGN), because in these cases the central emission in the $\rm{H\alpha}$ images 
does not trace recent star formation but the ionization of the gas caused by the AGN.  To overcome this problem, we used
the nuclear spectra from Gavazzi et al. (2011) to identify galaxies with an AGN.
Then, the central region was masked in these images to exclude the effects of the AGN emission. 
The standard radius for this mask was set to be 10 arcsec for nearby galaxies in the Local Supercluster 
and 2.5 arcsec for galaxies belonging to the Coma Supercluster\footnote{ The Coma Supercluster is five times farther away
than Virgo, therefore the masking radius should be 2 arcsec. We used 2.5 arcsec ($\sim 1.5 \times \rm{<PSF~FWHM>}$) 
to take into account the average seeing conditions of our images.
For extended low ionization nuclear emission regions (see Keel 1983; Hameed \& Devereux 1999)
which, in most cases are ionized by post-asymptotic giant branch (PAGB) stars (Stasi{\'n}ska et al. 2008), we used a masking radius
twice as large, i.e., 20 arcsec at the distance of Virgo and 5 arcsec at the distance of Coma.}.

We also excluded from the analysis of the CAS parameters 107 galaxies that are either undetected in $\rm{H\alpha}$ or that show only 
nuclear emission associated to an AGN. In the end, structural parameters were measured on  $r$-band images for 742/798 
galaxies and on H$\alpha$ images for 635/798 galaxies. 

\subsection{Concentration index} \label{Concentration}
Concentration is defined in slightly different ways by different authors, but it basically 
measures the ratio of light within a circular or elliptical inner aperture to the 
light within an outer aperture. The CAS system adopts the Bershady et al. (2000) 
definition as the ratio of the circular radii containing 20\% and 80\% of the ``total flux'':

\begin{equation}
C=5\times log \left(\frac{r_{80}}{r_{20}}\right),
\label{ewconcentration}
\end{equation}
where $r_{80}$ and $r_{20}$ are the circular apertures containing 80\% and 20\% of the total flux. 
As a definition of the total flux,  Conselice (2003) used the flux contained within 1.5 times the 
Petrosian radius\footnote{The Petrosian radius is the radius $r_p$ at which the ratio of the 
surface brightness at $r_p$ to the mean surface brightness within $r_p$ is equal to a fixed value, 
typically set to 0.2 (Petrosian 1976).} $r_p$ ($r_{80}$ and $r_{20}$ are considered only 
if they are larger than 1.5 times the seeing, as suggested by Arribas et al. (2012), otherwise
$C$ is not given).
For the measurement of $C$  the galaxy's center coincides with the center that minimizes the 
asymmetry (see below).

Yamauchi et al. (2005) pointed out that the concentration index can be overestimated for edge-on 
galaxies when circular apertures are used.
\begin{figure} [h]
 \centering
\includegraphics[scale = 0.65]{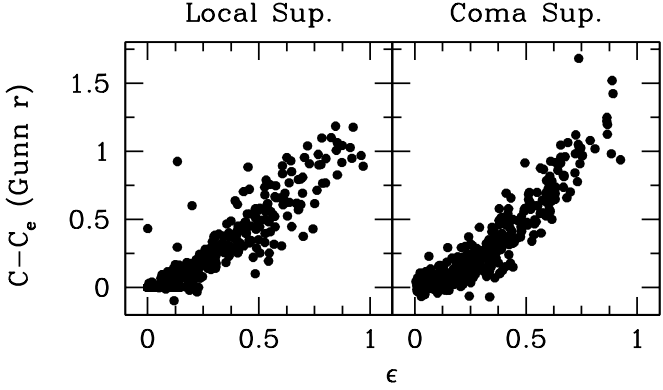}
\caption{Difference between the concentration index in the $r$-band computed in elliptical 
and circular apertures as a function of $\epsilon = \log_{10}(a/b)$ (de Vaucouleurs et al. 1991) for the Local Supercluster (left) and the Coma 
Supercluster (right) samples. }
\label{Cellcirc}
\end{figure} 
We performed multi-aperture photometry both in circular and elliptical apertures and we found a strong correlation 
between the axis ratio and the difference between the 
concentration index computed in circular ($C$) and elliptical ($C_e$) apertures, as showed in Fig. \ref{Cellcirc}.  
In other words, edge-on spirals have an abnormally high concentration of light if defined with circular 
apertures, and the morphological classifications based on $C$ will suffer from significant contamination. 
Using elliptical apertures remedies this problem, therefore our CAS results are based on the concentration index $C_e$.

\subsection{Asymmetry index}
The asymmetry parameter $A$ quantifies the degree to which the light of a galaxy is rotationally symmetric. 
$A$ is defined as the difference in flux between the original image and the one rotated by $180^\circ$ from its center,
normalized to the flux in the original image (Abraham et al. 1996; Conselice et al. 2000):
\begin{equation}
A=\frac{\sum_{i,j}|I(i,j)-R_{180}(i,j)|}{\sum_{i,j}|I(i,j)|},
\end{equation}
where $I$ is the original  image and $R_{180}$ is the image rotated by  $180^\circ$.  
$A$ is summed over all pixels ($i,j$) within 1.5 $r_p$ of the galaxy center.
  
The computation of the asymmetry is not robust with respect to the assumed photometric center,
since a positional difference of  1 arcsec  produces substantially different asymmetry values. 
To overcome this centering problem, we defined the rotation center as the position that yields 
a minimum value of the asymmetry parameter.  It should be noted that this issue 
especially affects the asymmetry parameter but, following Conselice (2003), Lotz et al. (2004), 
and Vikram et al. (2010), we used this centering method also compute the two other CAS parameters.

Operationally, we made an initial guess starting from the photometric center and then iteratively computed the 
asymmetry in a $5 \times 5$ pixel grid. Unlike those of other authors (Conselice 2003; Vikram et al. 2010), our grid steps 
are integer pixels Cotini et al. (2013) claim that the use of fractional pixels 
introduces a non-negligible smoothing that affects the CAS parameters.
The Petrosian radius was recomputed after each step, until the asymmetry reached a minimum. 
The asymmetry computed in this way can be overestimated if the background contains structure (i.e., 
haloes of bright stars or poor flat fielding). These effects were corrected for by simultaneously performing 
the asymmetry measurement on both the source and a neighboring blank area of the image. 
Taking into account these noise corrections, the equation used to compute the asymmetry is
\begin{equation}
A=\frac{\sum_{i,j}|I(i,j)-R_{180}(i,j)|}{\sum_{i,j}|I(i,j)|}-\frac{\sum_{i,j}|N(i,j)-NR_{180}(i,j)|}{\sum_{i,j}|I(i,j)|},
\end{equation}
where $N$ is the background image and $NR$ is the rotated background image.

\subsection{Clumpiness index}
The clumpiness parameter $S$ has been developed by Conselice (2003), inspired by the work 
of Takamiya (1999), to quantify the amount of light in small-scale structures. The clumpiness is defined 
as the ratio of the amount of light contained in high-frequency structures to the total amount of light in 
the galaxy. For ellipticals this ratio should be, and generally is, near zero.

First, the original galaxy effective resolution was reduced to create a 
blurred image ($B$). This was done by smoothing the galaxy with a boxcar filter of width $\sigma$. Although in principle 
any scale can be used to measure various frequency components in a galaxy, we followed Conselice (2003) 
and Lotz et al. (2004) setting a smoothing filter size $\sigma(I) = 1/5~\times~1.5~\times~r_p = 0.3~\times~r_p$. 
The effect of the blurring is to create an image whose high-frequency structures have been washed out. 
If a galaxy only consists of concentrated but structureless light (like a BCD, for example) it  
has a very low clumpiness. The blurred image was then subtracted from the original image ($I$).
The clumpiness parameter $S$ is defined as 
the flux in the residual image normalized to the flux in the original image. 

The central pixels within a circular aperture equal to the smoothing length $0.3 ~ r_p$ were excluded from the sum because 
the central parts of the galaxies are often unresolved and contain significant high-frequency structure due to finite sampling (Conselice 2003). 
Following the definition of Conselice, we also forced to zero any negative pixel in the residual map before 
computing $S$. This was achieved by substituting the original image with the one ($I_{max}$) obtained taking the highest 
value between $I$ and $B$ for each pixel.

Even the clumpiness can be overestimated by noise effects caused by the graininess of the background. 
We corrected this source of noise in the same way as for the asymmetry, and the final equation used to compute 
clumpiness becomes
\begin{equation}
S=10\frac{\sum_{i,j}[I_{\rm max}(i,j)-B(i,j)]}{\sum_{i,j}[I_{\rm max}(i,j)]}-10\frac{\sum_{i,j}[N_{\rm max}(i,j)-NB(i,j)]}{\sum_{i,j}[I_{\rm max}(i,j)]},
\end{equation}
where $N_{\rm max}$ is the maximum background image and $NB$ is the blurred background image.

\subsection{Comparison with the literature}

The code developed for the morphological analysis was tested on 
18  Local Supercluster galaxies in common with  Frei et al. (1996) that have also been analyzed  
by Conselice (2003). These are all giant, nearby NGC galaxies with $M_B < -20~mag_{\rm vega}$. 
\begin{figure*}[tb]
\begin{center}
\includegraphics[width=0.85\textwidth]{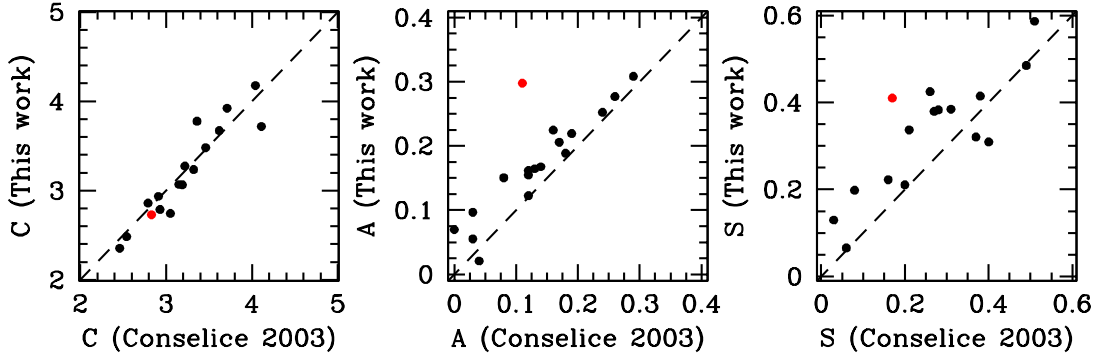}
\caption{Comparison between CAS parameters computed in this work and those given by Conselice (2003). The dashed
lines mark the 1:1 relation. The red point is the VCC 1987 (NGC 4654).}
\label{comparisonCONS}
\end{center}
\end{figure*}

Our estimates of the CAS parameters for this subsample are listed in Table \ref{comparisonCONSt}
along with details on the telescope, the pixel scale and the seeing conditions (Full width half maximum 
of the point spread function, PSF FWHM) of the images. The Frei et al. images have been taken with similar 
$r$ filters, but in slightly poorer seeing conditions (PSF FWHM $\sim~ 2" - 5"$).

The comparison between the three CAS parameters computed using our in-house codes and those given in 
Conselice (2003) are shown in Fig. \ref{comparisonCONS}.
The discrepant red point in the $A$ and $S$ comparisons is  VCC1987 (NGC4654), a giant Sc galaxy in the Virgo cluster. This galaxy 
appears to be far more asymmetric and clumpy in our analysis than in Conselice (2003). We have several reasons to believe that this galaxy
is in fact asymmetric. First, in the analysis of Conselice, this galaxy seems to be less asymmetric and clumpy 
than the remaining Sc galaxies in the sample. Furthermore, it is known that this galaxy is strongly 
asymmetric in the stellar disk, in atomic gas (HI) (Chung et al. 2007), in molecular gas ($\rm{H}_2$) 
(Cayatte et al. 1990; Chung et al. 2009), and also in the $\rm{H\alpha}$ velocity field (Chemin et al. 2006). 
For these reasons we excluded this galaxy from the comparison. 
\begin{table*}[tb]
\caption{Observational specifications and CAS parameters computed in this work for a set of 18 galaxies in common with
 Conselice (2003).}
\begin{center}
\begin{tabular}{c c c c c c c c c c}
\hline
\hline
NGC & Telescope\tablefootmark{a} & Plate Scale  & Seeing & ${C}_r$ &${A}_r$  & ${S}_r$ &${C}_r$  &${A}_r$  & ${S}_r$   \\
    &  &  ($\rm{arcsec~pixel}^{-1}$) & (arcsec) &\multicolumn{3}{c}{This Work} & \multicolumn{3}{c}{Conselice (2003)} \\
\hline
3596 & SPM  & 0.31 & 1.39 & 2.79 & 0.17 & 0.34  & 2.93 & 0.14  & 0.21 \\ 
4189 & SPM  & 0.31 & 2.20 & 2.35 & 0.28 & 0.32  & 2.46 & 0.26  & 0.37 \\ 
4192 & OHP  & 0.69 & 2.62 & 3.67 & 0.22 & 0.64  & 3.62 & 0.19  & 0.36 \\ 
4216 & OHP  & 0.69 & 3.51 & 5.16 & 0.21 & 0.41  & 5.08 & 0.17  & 0.38 \\ 
4254 & OHP  & 0.69 & 1.93 & 3.24 & 0.31 & 0.48  & 3.32 & 0.29  & 0.49 \\ 
4303 & CA   & 0.41 & 1.59 & 2.86 & 0.25 & 0.59  & 2.79 & 0.24  & 0.51 \\ 
4321 & OHP  & 0.69 & 2.10 & 2.94 & 0.16 & 0.38  & 2.91 & 0.12  & 0.28 \\ 
4501 & OHP  & 0.69 & 2.41 & 3.27 & 0.16 & 0.21  & 3.22 & 0.13  & 0.20 \\ 
4535 & OHP  & 0.69 & 2.65 & 2.48 & 0.12 & 0.31  & 2.54 & 0.12  & 0.40 \\ 
4548 & CA   & 0.41 & 1.64 & 3.92 & 0.07 & 0.13  & 3.71 & 0.00  & 0.03 \\ 
4569 & OHP  & 0.69 & 3.86 & 3.78 & 0.15 & 0.22  & 3.36 & 0.08  & 0.16 \\ 
4579 & CA   & 0.41 & 1.64 & 4.18 & 0.06 & 0.07  & 4.04 & 0.03  & 0.06 \\ 
4654 & CA   & 0.41 & 1.68 & 2.73 & 0.30 & 0.41  & 2.83 & 0.11  & 0.17 \\ 
4689 & CA   & 0.41 & 1.72 & 2.74 & 0.01 & 0.20  & 3.05 & 0.03  & 0.08 \\ 
5248 & SPM  & 0.31 & 1.79 & 3.48 & 0.22 & 0.42  & 3.46 & 0.16  & 0.26 \\ 
5364 & SPM  & 0.31 & 1.39 & 3.07 & 0.15 & 0.38  & 3.15 & 0.12  & 0.27 \\ 
5669 & SPM  & 0.31 & 1.44 & 3.06 & 0.19 & 0.38  & 3.19 & 0.18  & 0.31 \\ 
5701 & SPM  & 0.31 & 1.73 & 3.72 & 0.02 & -0.01 & 4.11 & 0.04  & 0.06 \\ 
\hline
\end{tabular}
\tablefoot{
\tablefoottext{a}{SPM = San Pedro Martir 2.1m; 
 CA = Calar Alto 3.5m; OHP = Observatoire de Haute Provence 1.2m.}}
\label{comparisonCONSt}
\end{center}
\end{table*}

The agreement for the other points is remarkable. 
We found that the average deviation between our values and the published ones for concentration, asymmetry, and clumpiness 
are $ -0.02 \pm 0.18, ~ 0.03 \pm 0.03, ~\rm{and}~ 0.06 \pm 0.09$. Similar dispersions were seen in
previous comparisons. The dispersion between the CAS parameters reported by Vikram et al. (2010) from that of
Conselice (2003) are $ -0.11 \pm 0.14, ~ 0.00 \pm 0.04, ~\rm{and}~ 0.06 \pm 0.09$ using the whole Frei et al. (1996) sample.
The agreement in these results  gives an idea of the robustness of our software,
but also of the stability of the CAS parameters when using different source images. We verified that convolving
our images to a common PSF (with FWHM $= 3"$) does not significantly change the comparison with published values. 
Any resolution effect, if present, is negligible due to the small seeing differences between our images and those by 
Frei et al. (1996) and to the limited statistics. Only by comparing the results from our whole sample with those of Conselice (2003) 
a small systematic difference can be seen and ascribed to a resolution effect (see Sect. \ref{resultCASr}).
Indeed, Lotz et al. (2004) found that C, A, and S parameters are reliable within $\pm$ 0.3, 0.1, and 0.2 when
the S/N per pixel, within $1.5~\times~r_p$, is $<S/N>~\ge~5$.

\section{Size measurements} \label{sizes}

\begin{figure*}[!t]
\begin{center}
\begin{tabular}{ c c c}

Gunn $r$ & $\rm{H\alpha}$ NET &  $\rm{H\alpha}$ EW \\
 \includegraphics[scale=0.30]{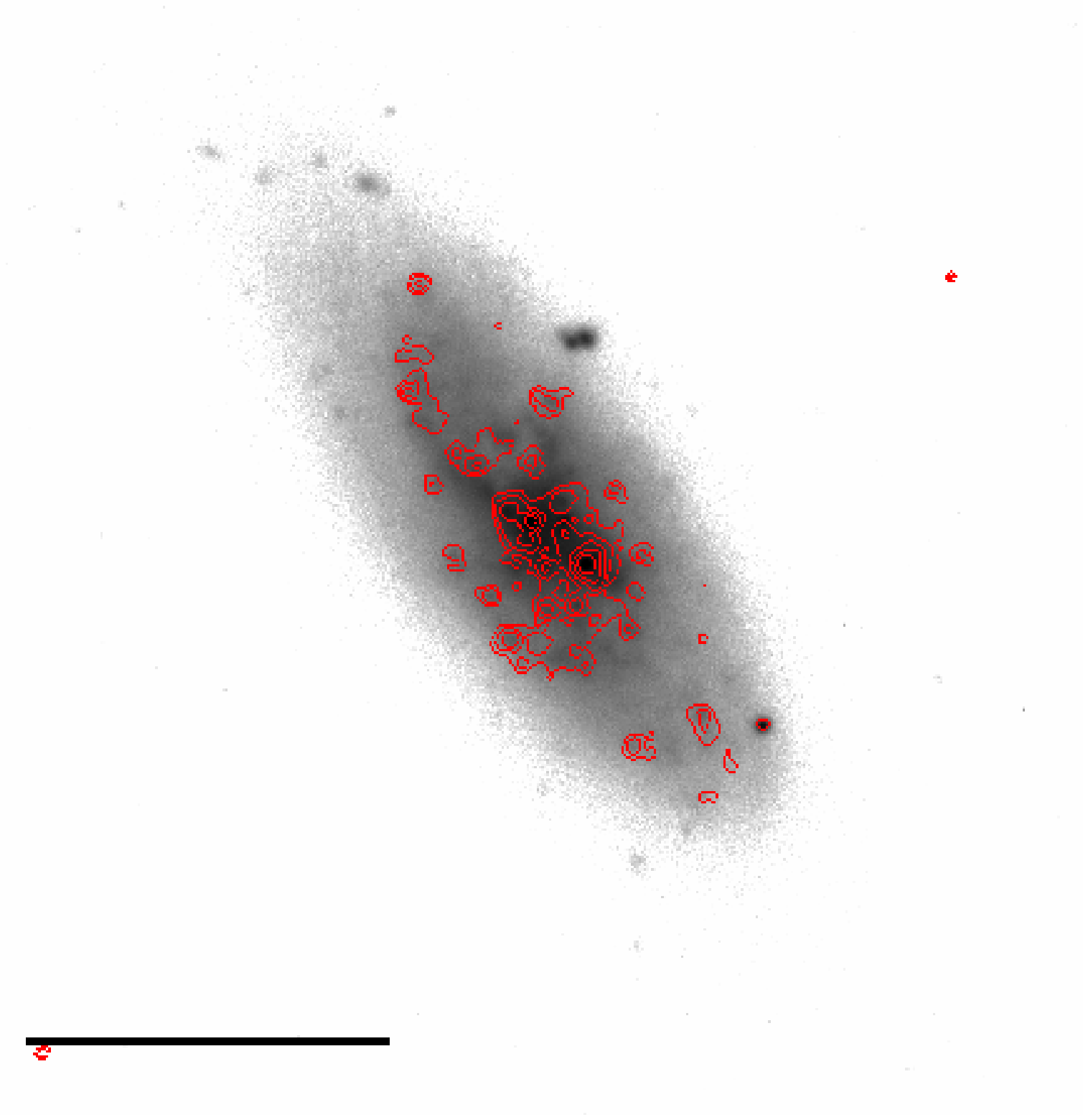} & \includegraphics[scale=0.30]{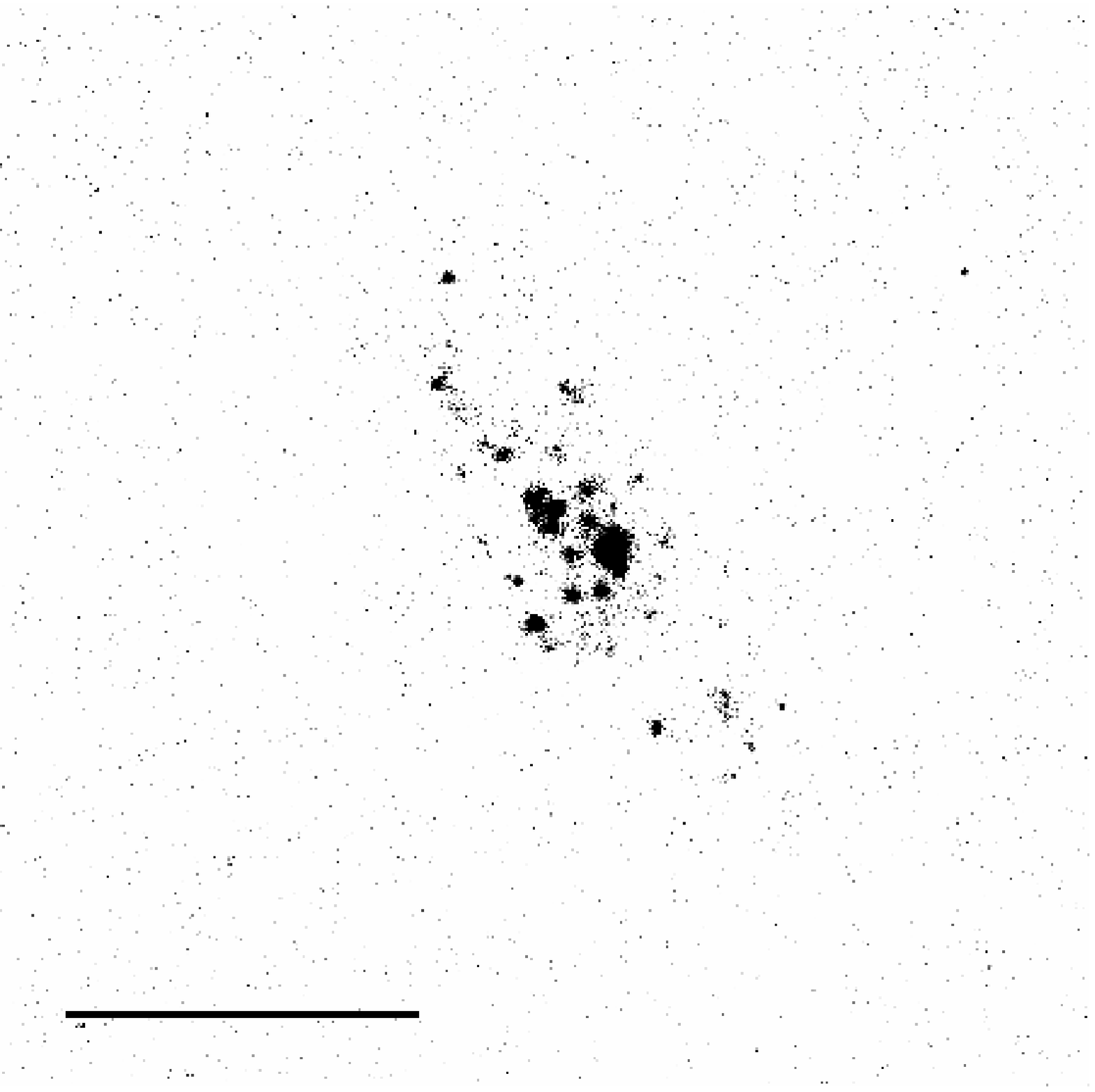} & \includegraphics[scale=0.30]{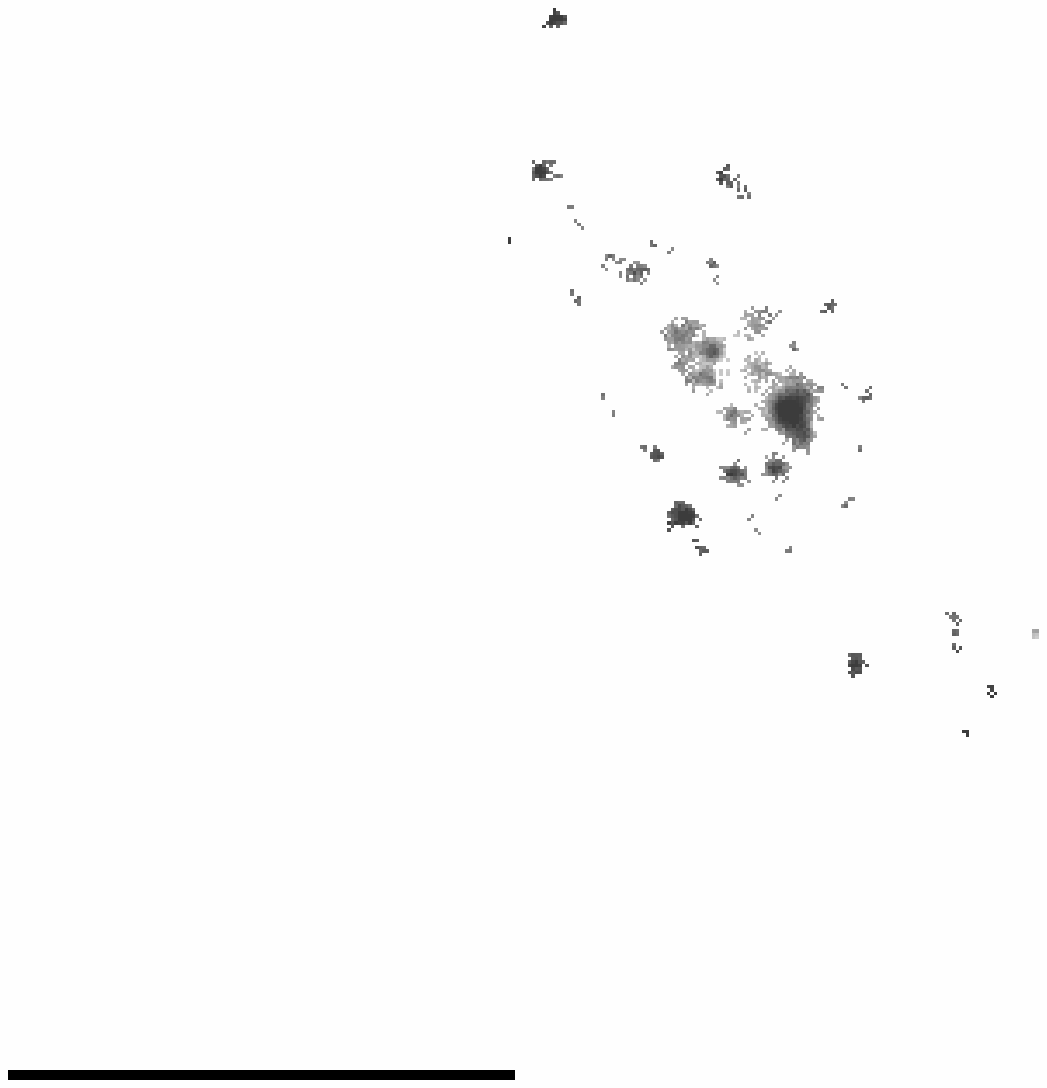} \\
 \includegraphics[scale=0.30]{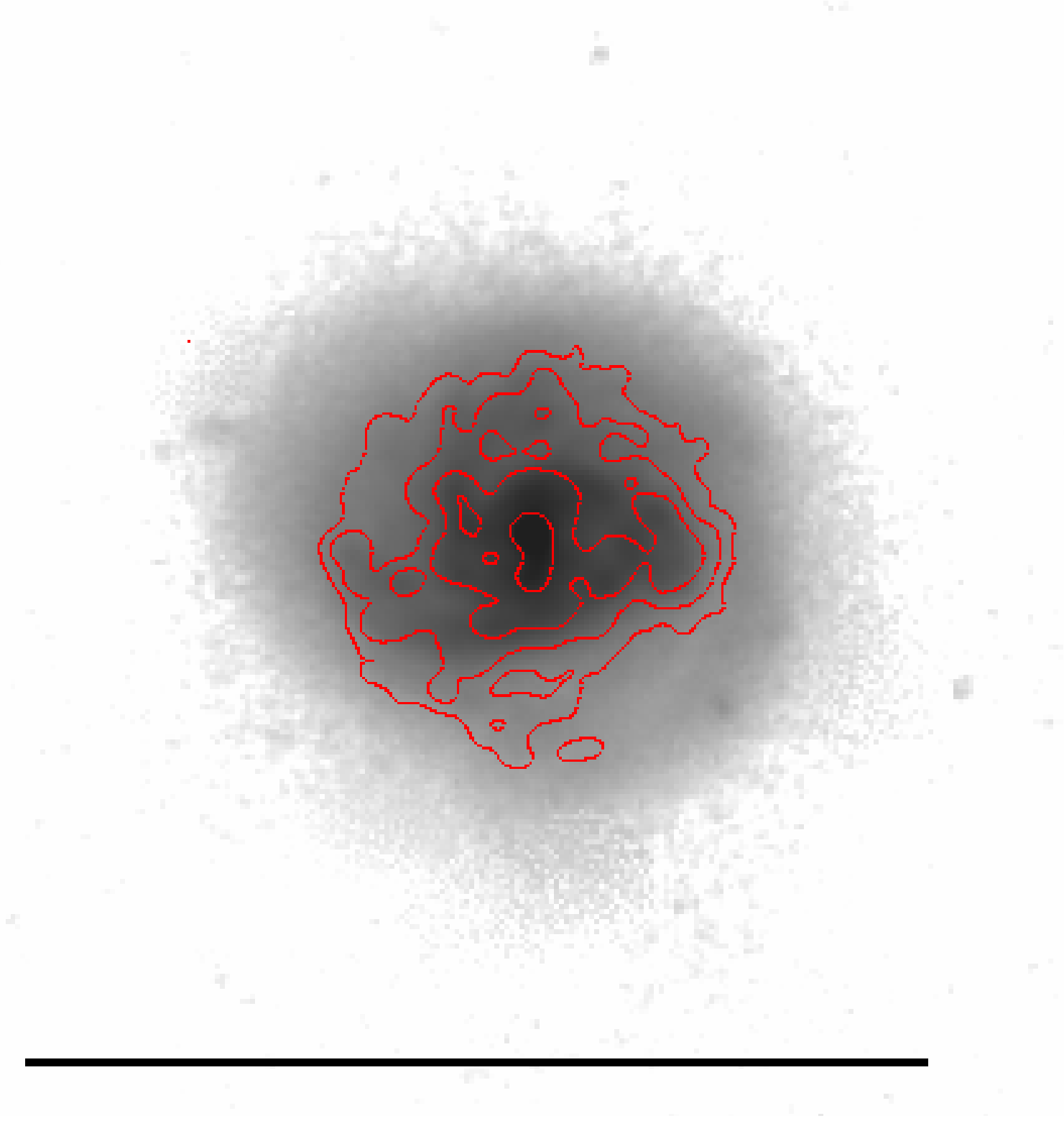} & \includegraphics[scale=0.30]{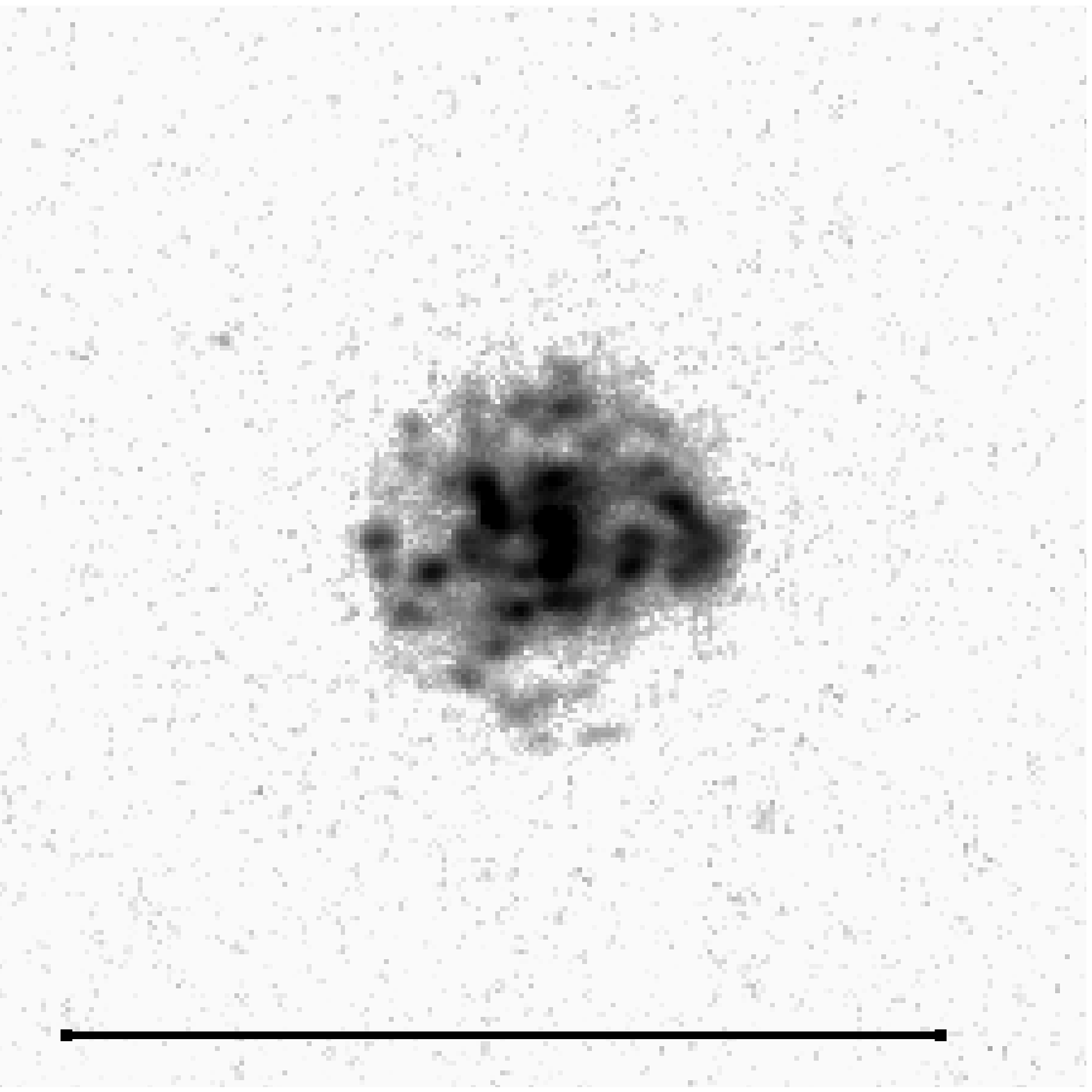} & \includegraphics[scale=0.30]{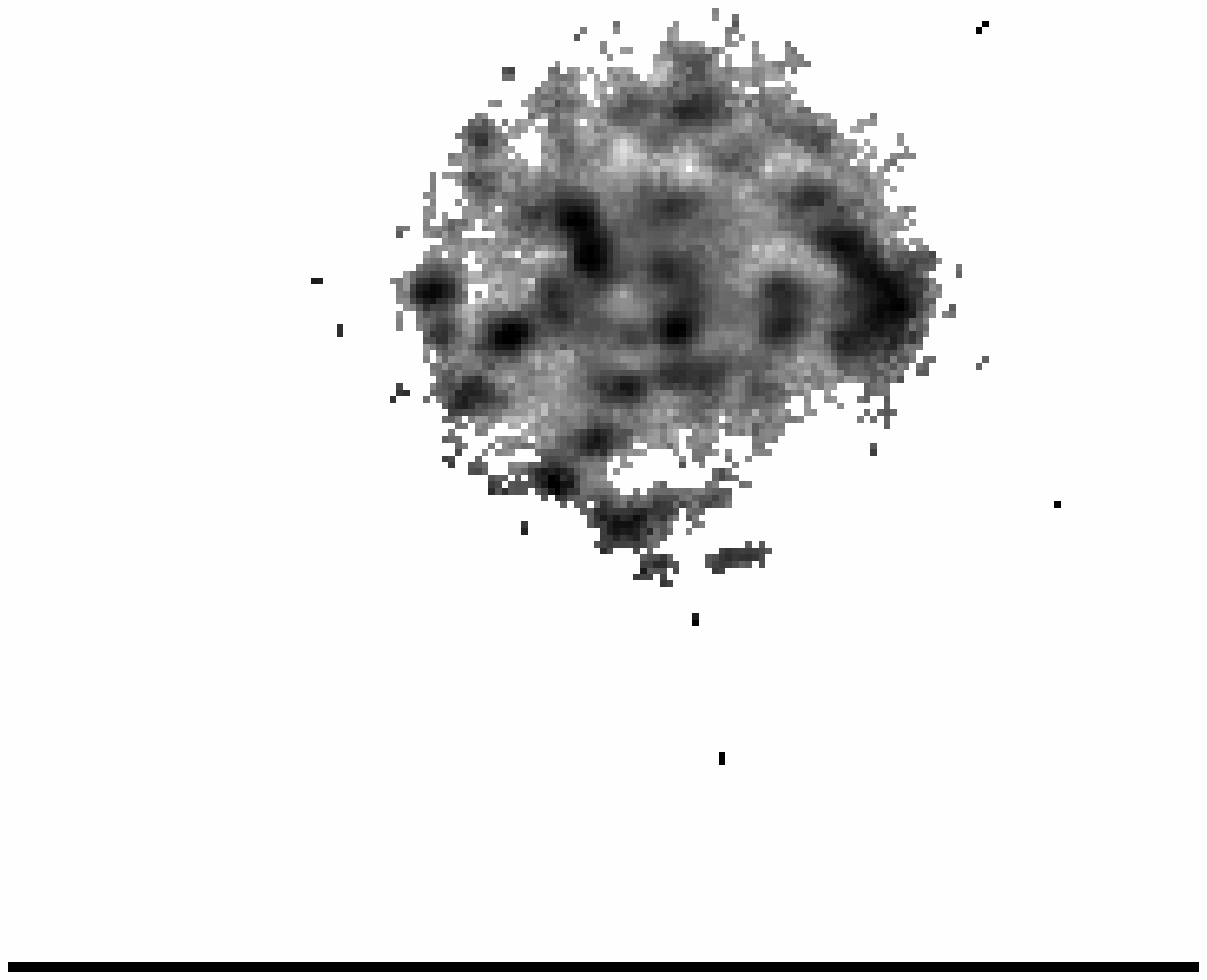} \\
 \includegraphics[scale=0.30]{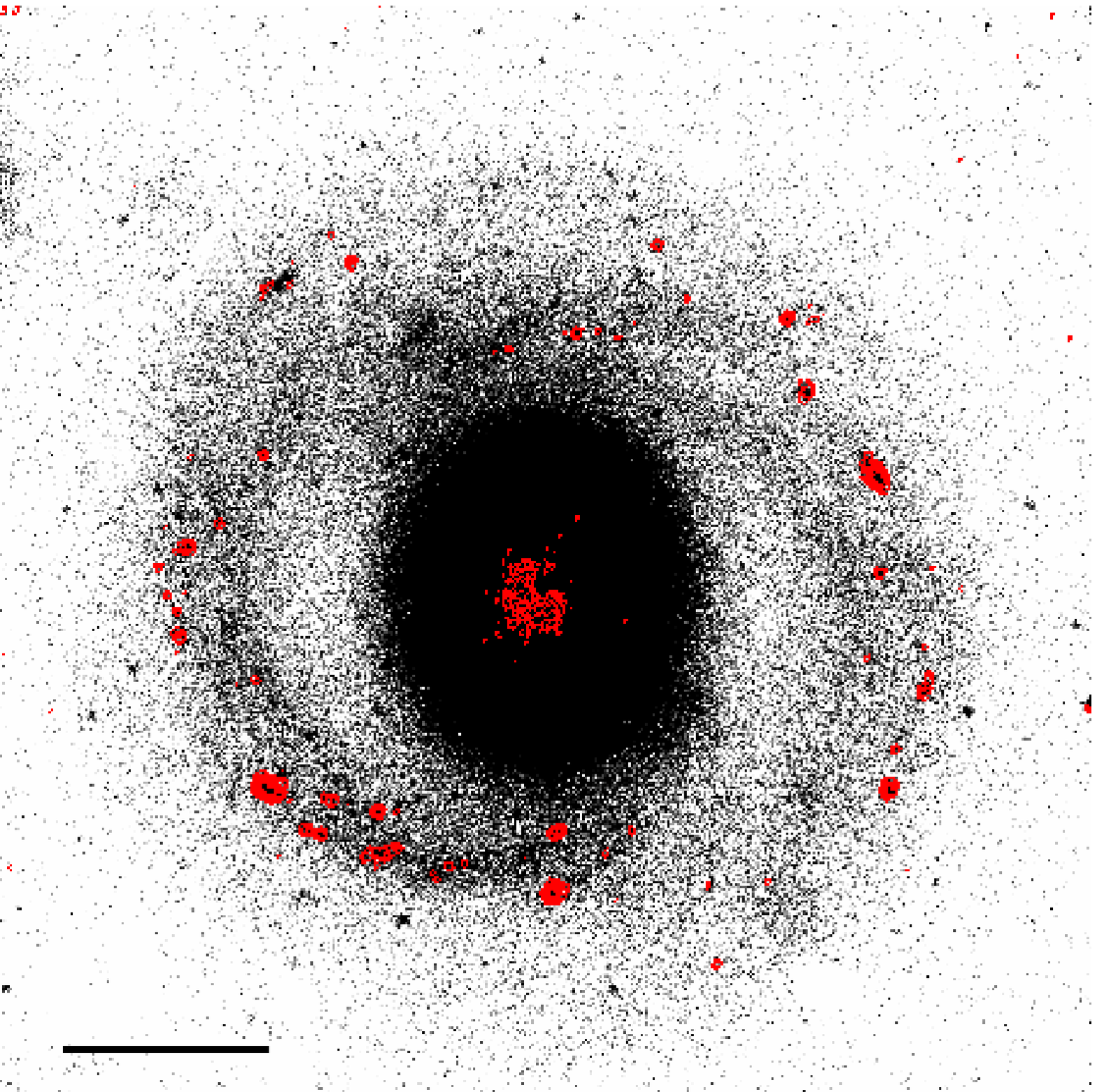} & \includegraphics[scale=0.30]{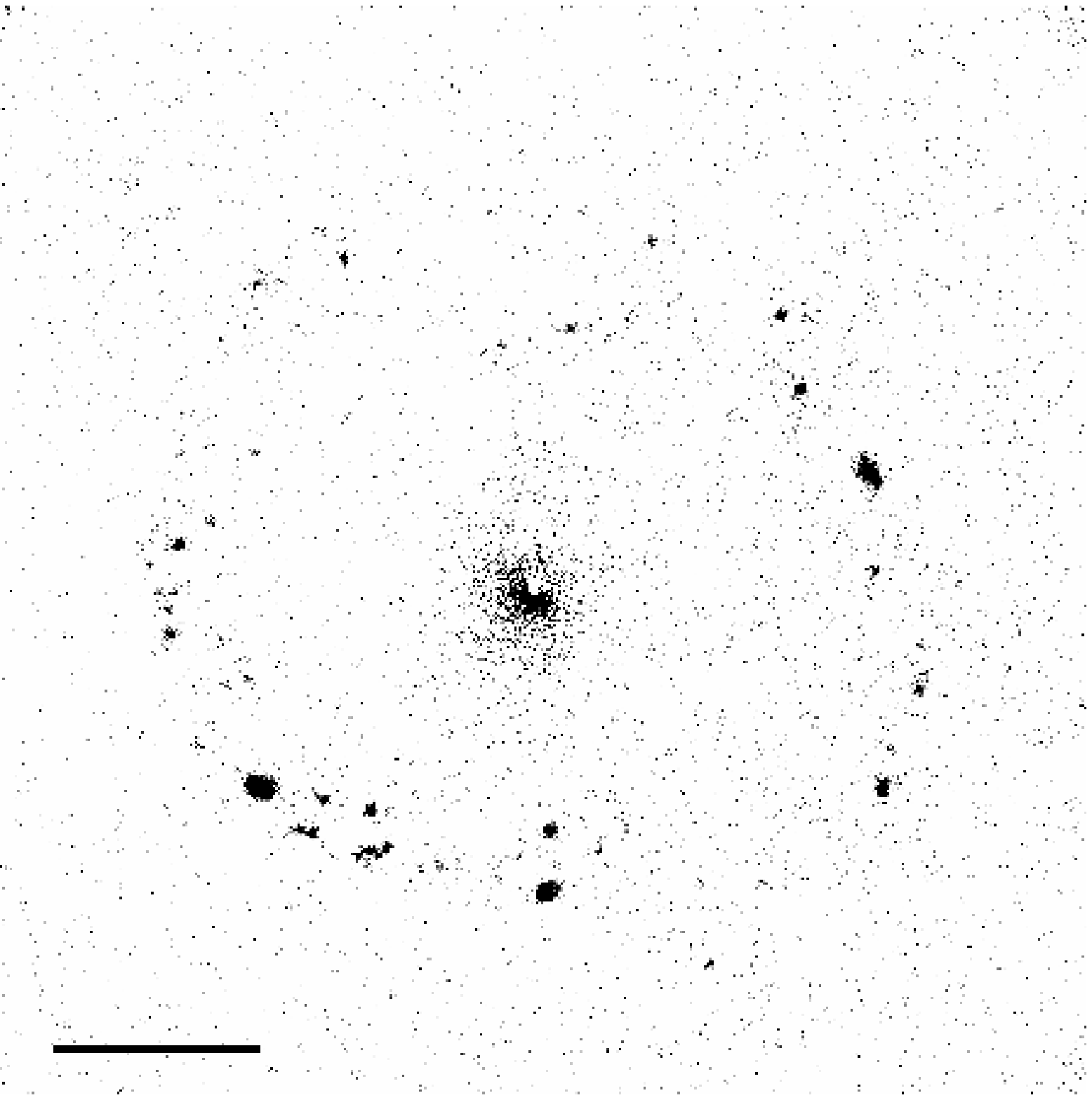} & \includegraphics[scale=0.30]{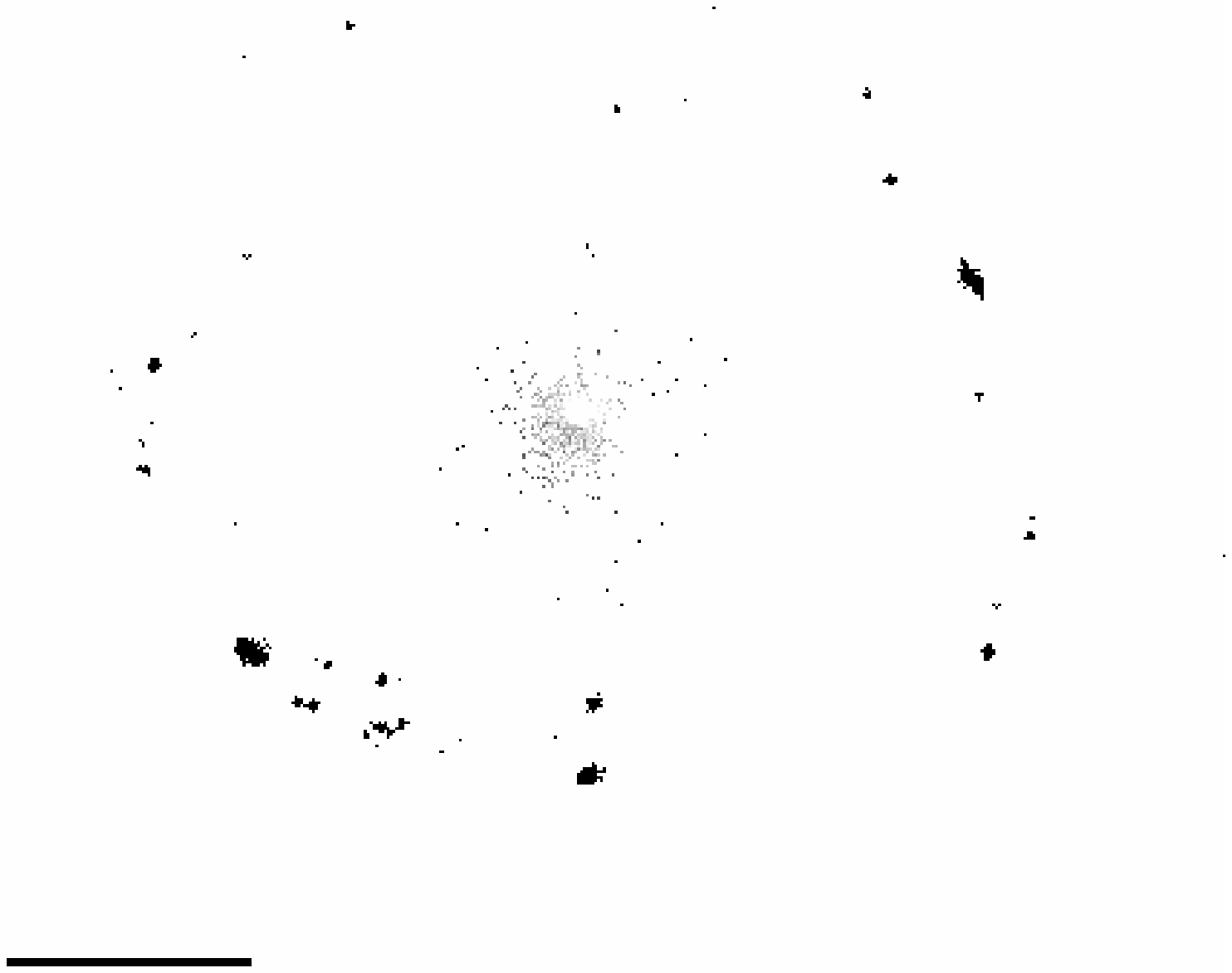} \\
 
\end{tabular}
\caption{Gunn $r$ band (left), $\rm{H\alpha}$ (center), and EW (right) images of three galaxies spanning the full range of 
the $EW/r$ parameter: from $EW/r = 0.55$ (AGC 7874: top), $EW/r =1.06$ (CGCG 160-098: middle), to $EW/r = 3.87$ (AGC 7874: bottom). 
The red contours are drawn from the $\rm{H\alpha}$ map, and the black line marks the 1 arcmin scale.}
\label{imaH}
\end{center}
\end{figure*} 
To investigate on which spatial scale galaxies are growing, we parameterized the
spatial distribution of the old and the young stellar populations. 
The stellar mass of a galaxy was traced using the Gunn $r$ band and the
star-formation was traced by the $\rm H\alpha$ line emission. Some uncertainties that affect the size 
measurements are to be considered. First, $\rm{H\alpha}$ $\lambda6563$ and the [NII] doublet  
$\lambda\lambda~6548,6584$ are not spectrally resolved by means of narrow band interferential filters. 
We do know the line ratios for a large number of objects from integrated drift-scan spectroscopy (Gavazzi et al. 2004; 
Boselli et al 2012), although they are of limited use for our purpose, since line ratios almost certainly 
vary with the position in the galaxy (Genzel et al. 2008; Yuan et al. 2011). Therefore we did not apply any mean correction.
We tried to reduce this problem by masking the AGN emission (dominated by [NII]), where present, using nuclear 
spectra (see below).
A second systematic uncertainty is dust attenuation. Differential extinction could affect the spatial 
distribution of the $\rm H\alpha$ emission (Boissier et al. 2004, 2007), and thus the size measurements.  
Finally, the Gunn $r$ filter contains the $\rm H\alpha$ (and [NII]) lines, which introduces a correlation
between the two images. Nevertheless ,this effect is small because the lines' contribution to the stellar 
continuum images is lower than $10\%$ of the total flux (Paper I).

\subsection{Effective radii}
The effective radius (or half-light radius) is a parameterization of a galaxy's light 
distribution. It is defined as the radius containing half of the total light in a specific band. 
The standard method for determining half-light radii of galaxies is to fit two-dimensional 
models to the galaxy light profiles (e.g. using GALFIT, Peng et al. 2002).
However, this parametric method does not provide meaningful results for the $\rm{H\alpha}$ images 
because they are typically clumpy and often have asymmetric or centrally depressed light profiles 
(Shapley 2011). For this reason we followed the method suggested by 
F{\"o}rster Schreiber et al. (2009), Arribas et al. (2012), and Nelson et al. (2012). We measured effective radii for 
both the $\rm{H\alpha}$ images and the optical $r$-band continuum using integrated curves of growth (CoG). 
Using this method, the effective radii depend on the limiting sensitivity of the images.
However, we compared the effective radii measured from CoGs with those obtained fitting the radial light profiles
by Cortese et al. (2012) on a sample not presented here (Fossati et al. in prep.), and found only small differences
($\sim10\%$) between the two methods.
Another uncertainty can arise if the limiting sensitivity of narrow-band $\rm{H\alpha}$ images and that of 
the $r$ broad-band are different. The sensitivities in those two bands are comparable if the exposure times are
in the same ratio as the filter bandpasses. We have a range of these exposure time ratios spanning from $60\%$
to $100\%$.  We verified that using the galaxies observed in the correct ratio of the exposure times 
and simulating $40\%$ shorter exposures in the narrow-band filter, the derived effective radius $r_e(\rm{H\alpha})$ 
typically varies by only $\sim 5\%$. We conclude that even in the worst conditions this uncertainty is negligible.  

The CoG procedure is similar to the one used to compute the concentration index, with some minor 
differences. The center of each galaxy here is the photometric center and not the symmetry center.
To achieve a better comparison between the two different bands, the $\rm{H\alpha}$ center was forced to 
coincide with the $r$-band center. Furthermore, the elliptical apertures used for $\rm{H\alpha}$ photometry were 
defined by $r$-band isophotes.
The effective radii were defined as the radii containing 50\% of the maximum curve of growth value. 
The effective radii were considered reliable only if they exceeded 1.5 times the seeing in both bands 
(Arribas et al. 2012). For AGN the central region of the $\rm{H\alpha}$ image was masked 
before computing of the $\rm{H\alpha}$ effective radius.

\subsection{$EW/r$ parameter}   
Visual inspection of the $\rm{H\alpha}$ images reveals that the bulge of the early type spirals 
often shows diffuse line emission. This can be due to ionization from AGNs, or from red and 
old PAGB stars (Stasi{\'n}ska et al. 2008) or a technical artifact due to improper normalization 
of the OFF-band images. Indeed, as pointed out by Spector et al. (2012), the normalization
coefficient of the $r$- on the narrow-band image strongly depends on the color of the galaxy. 
We calibrated our normalization for blue galaxies ($<g-r> \; = 0.4$) because our sample is 
dominated by blue dwarfs. The issue arises for the bulges of massive spirals that can be up to 1 mag 
redder. In these cases the underlying continuum may be under-subtracted (up to a factor of $\sim 10\%$),
leaving a residual in the $\rm{H\alpha}$ images that is completely unrelated to ongoing star formation.

To overcome this technical problem and the additional uncertainties due to central AGN-like emission, 
we defined a new parameter, called $EW/r$, as the ratio between 
the effective radii computed  in the equivalent width of the $\rm{H\alpha}$ (EW) map and in the $r$-band image.
By dividing the $\rm{H\alpha}$ by the $r$-band continuum images 
\footnote{To achieve a real EW, the pixel values should be 
multiplied by the width of the narrow band filter used for the line observations, 
but because we do not require absolute calibration to compute effective radii 
(they are normalized to the maximum value of the curve of growth) we multiplied the pixel 
values by the arbitrary value of 100. This was to avoid systematic errors 
due to the poor handling of small pixel values by Funtools.}, 
we obtain a map of the EW that is defined as the strength of a line
normalized to the underlying continuum.
The signal of emitting HII regions with a shallow continuum is enhanced in this image, and on the
other hand the signal in the regions dominated by the old stellar populations, like in the bulges, is
quenched. 

Another important correction needs to be performed before dividing the $\rm{H\alpha}$ and $r$ images. 
Since the background of both images is zero on average with a Gaussian distribution, dividing 
the two images can produce spurious high values in pixels with a low 
$r$-band flux. Using the {\sc IRAF} task \emph{imreplace} we replaced all pixel values from 
$-\infty$ to $+3\sigma_{sky}$ with zero. Then, by setting the {\sc IRAF} task \emph{imarith} to place a 
0 when trying to divide by zero, we obtained an EW image with zero background below 
the sensitivity limit\footnote{The few isolated pixels greater than the $+3\sigma_{sky}$ limit have 
been replaced with a 0 using a Python script.}.
It should be noted that setting such a sensitivity limit could lead to an
underestimate of $EW/r$. 
Nonetheless, we verified that lowering the limit to $+2\sigma_{sky}$
increases $EW/r$ by $\sim10\%$, but also increases the probability that noisy
regions in the background around the galaxy are included in the galaxy flux.
Since the EW image translates into a map of the specific star formation 
rate\footnote{ The specific star formation rate (SSFR) is the star formation rate normalized 
to the stellar mass of a galaxy.}, this $EW/r$ parameter traces the position of newly born 
stars weighted on the distribution of the old stars, relative to the position of the old stars.
It is computed as follows:
\begin{equation}
EW/r = \frac{r_e(\rm{EW_{\rm{H\alpha}}})}{r_e(\rm{Gunn}~r)}.
\label{hierarchyeq}
\end{equation}
If this parameter is greater than 1, new stars are born outside 
the already constituted assembly of old stars. On the other hand, galaxies whose 
$EW/r$ is smaller than 1 are building up their central concentration. Figure \ref{imaH} shows the 
$r$-band,  $\rm{H\alpha}$, and EW images of three galaxies spanning the full range of the $EW/r$ parameter.

\section{Results} \label{resultsect}

\subsection{The CAS parameter space in the $r$ band} \label{resultCASr}

When investigating the three-dimensional space of CAS parameters computed in the Gunn $r$ band
we must remove edge-on galaxies from our sample because clumpiness (and to a lesser extent also asymmetry)
is affected by inclination. This is because the presence of a projected central dust line completely 
changes the apparent morphology of edge-on galaxies as defined in the CAS relative to nearly face-on objects.
To avoid this projection effect it is recommended to work with galaxies whose axial ratio is smaller than the disk oblateness.
Conselice (2003) studied the dependence of $S$ as a function of the axial ratio and, following his results, 
we set a rather conservative limit at $\epsilon < 0.4$, where $\epsilon = \rm{log_{10}}(a/b)$  as defined in
de Vaucouleurs et al. (1991). 
This limit corresponds to galaxies with an inclination lower than $66.5^\circ$ on the sky.
\begin{figure*}[t]
\begin{center}
\includegraphics[width=0.95\textwidth]{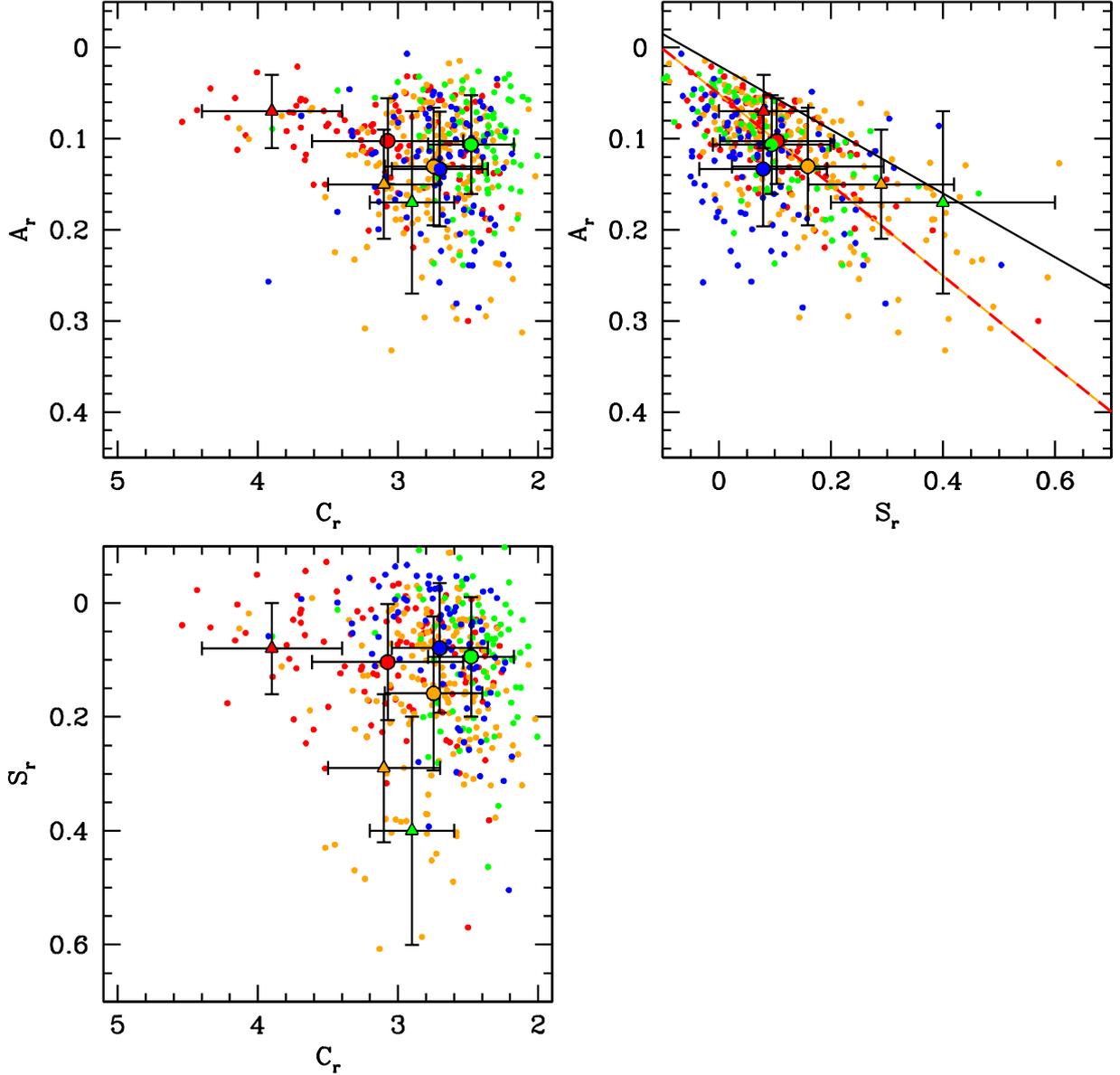}
\end{center}
\caption{Galaxy distribution (with inclinations lower than $66.5^\circ$ on the sky) in 
the space of CAS parameters measured in the $r$ band. The solid black line is the best fit of 
Conselice's (2003) data and the red-orange line is the best fit of our data restricted 
to the spirals (Sa-Sd) to be consistent with the Frei et al. (1996) selection used by 
Conselice. Points are color-coded according to the morphological type: red for Sa-Sb, orange for Sc-Sd, 
green for Sdm-Im, and blue for the BCDs.
The black contoured round points are the average values for each group of morphological type, and the black 
contoured triangles are the average values given by Conselice (2003) for the same group of types. Error bars
mark the $1\sigma$ standard deviation. }
\label{CASr}
\end{figure*}

\begin{figure*}[t]
\begin{center}
\includegraphics[width=0.95\textwidth]{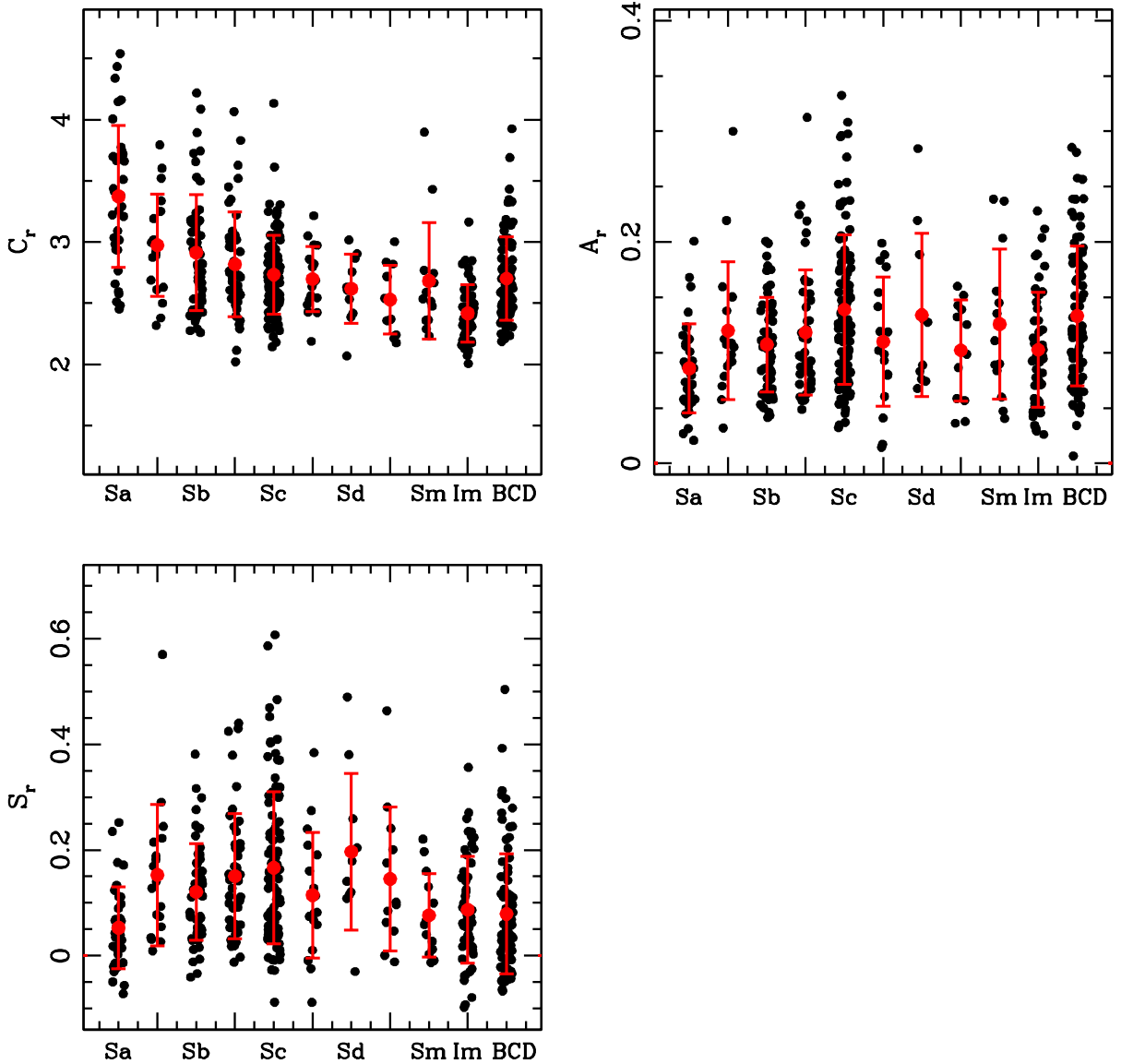}
\end{center}
\caption{CAS parameters in the $r$ band (for galaxies with an inclination lower than $66.5^\circ$ on the sky) as a function
of Hubble type. The red points are the average values for each type and the error bar marks the $1\sigma$ standard deviation.}
\label{CAStype}
\end{figure*}
Giant spirals are clearly separated in the $C_r$ vs $A_r$ panel (see Fig. \ref{CASr})
since the concentration index decreases along the Hubble sequence toward later types (and with decreasing stellar mass). 
The BCDs are as concentrated and asymmetric as the Sc-Sd, although they are dwarfs. 
The $A_r$ vs $S_r$ panel is the only one showing a linear correlation between 
the parameters, as already noted by Conselice (2003). The best bisector fit (Isobe 1990) of our data is
\begin{equation}
A_r = (0.48 \pm 0.02) \times S_r + (0.05 \pm 0.01).
\end{equation}
Both the slope and the intercept of the fit are higher than what was found by Conselice (2003), probably because of
the better resolution and seeing of our images compared to those of Frei et al. (1996). This is highlighted also by the fact that only 
a few points lie above Conselice's best fit (black solid line in Fig. \ref{CASr}). 
Finally, different Hubble types are, albeit with significant scatter, separated in the $S_r$ vs $C_r$ panel. 
The Sa-Sb giant spirals are the highest concentrated and their clumpiness is similar to that of the dwarf irregulars. 
The Sc-Sd intermediate-mass spirals are more clumpy but less concentrated than the Sa-Sb. 
The irregulars are the least concentrated but still clumpy, and the BCDs are as concentrated as 
the Sc-Sd, but are definitely not clumpy (see Fig \ref{CAStype}). 

\begin{table}
\begin{center}
\caption{Average values and $1\sigma$ rms of the CAS parameters computed in the Gunn $r$ band 
for galaxies with an inclination lower than $66.5^\circ$ on the sky in bins of Hubble type.}
\begin{tabular}{l c c c }
\hline
\hline
Type & $C_r$ & $A_r$ & $S_r$   \\
\hline
Sa - Sb      &  $ 3.1 \pm 0.5 $  & $0.10 \pm 0.04 $ & $ 0.10 \pm 0.09 $ \\   
Sc - Sd      &  $ 2.8 \pm 0.3 $  & $0.13 \pm 0.06 $ & $ 0.16 \pm 0.14 $ \\  
Sdm - Im     &  $ 2.5 \pm 0.3 $  & $0.11 \pm 0.05 $ & $ 0.09 \pm 0.10 $ \\  
BCD          &  $ 2.7 \pm 0.3 $  & $0.13 \pm 0.06 $ & $ 0.08 \pm 0.11 $ \\  
\hline                                                                                                                                                           
\end{tabular}                                              
\label{tab:CAStype}
\end{center}
\end{table}
Table \ref{tab:CAStype} gives the average values and the standard 
deviations of the CAS parameters in the same bins of type as reported by Conselice (2003). 
The agreement with previous studies is remarkable for spirals and for the concentration of the irregulars. 
We expect the mean values of $C$ in our work to be lower than those published by Conselice (2003) because we 
made use of elliptical apertures. We verified that this difference in the method increases the mean values
of $C$ by $\sim 0.1 - 0.3$.
Furthermore, we note that the sample used by Conselice (2003) consists of spiral galaxies 
more massive than those considered here ($M_* > 10^{9.5}M_\odot$). If we restrict our analysis
to the galaxies in common with Conselice (2003) (see Table \ref{comparisonCONSt}) and using circular apertures,
we obtain mean values of $C$ of $4.0 \pm 0.5$ and $3.0 \pm 0.3$ for Sa-Sb and Sc-Sd. 
Although the statistics is insufficient to rule out every other possible explanation, the results are now remarkably consistent.
Hern{\'a}ndez-Toledo et al. (2008) analyzed a sample of 539 isolated galaxies from the SDSS. 
The concentration index appears to be overestimated by $\sim 10\%$ in their work compared to ours because they use circular apertures 
(see Sect. \ref{Concentration}). They also found mean values for the $A$ and $S$ parameters in the $r$ band that are consistent with those 
in Conselice (2003) and in this work for Sa-Sb galaxies. 

\begin{figure*}[t]
\begin{center}
\includegraphics[width=0.95\textwidth]{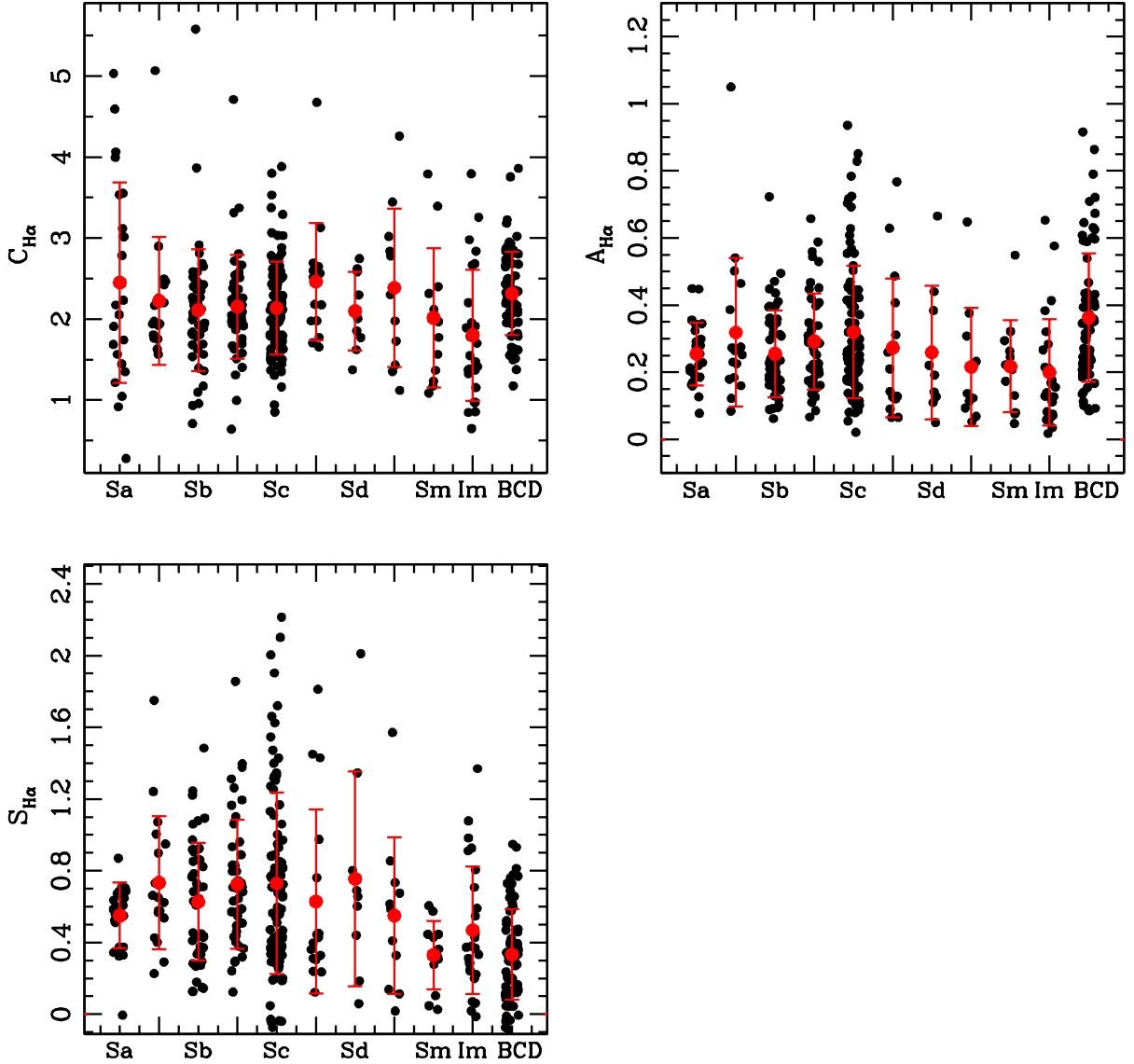}
\end{center}
\caption{CAS parameters in the $\rm H\alpha$ images (for galaxies with an inclination lower than $66.5^\circ$ on the sky) as a function
of Hubble type. The red points are the average values for each type and the error bar marks the $1\sigma$ standard deviation.}
\label{CAShatype}
\end{figure*} 

We did not find, in our sample that dwarf irregulars are more asymmetric and significantly
more clumpy than spirals, as found by Conselice (2003).
Conversely, we detected a decrement of these values going from spirals to the irregulars, the latter populating 
the region of the CAS space at $S \sim 0.1$. The same result ($S \sim 0.2$) was obtained by Hern{\'a}ndez-Toledo et al. (2008), 
although they did not consider late spirals (Sc-Sd) separately from dwarf irregulars.  
A possible explanation might be a different selection and definition of irregulars. Conselice (2003) used the dwarf irregular 
sample selected by van Zee (2000).  
This sample is composed of 58 galaxies selected from the UGC (Nilson 1973) catalog with $M_B > -18~mag_{\rm vega}$ without completeness in 
volume or surface brightness.
These galaxies are on average intrinsically brighter than the galaxies detected in HI by ALFALFA due to UGC incompleteness 
below $M_B \sim -16~mag_{\rm vega}$. Inspecting the images shown by van Zee (2000), we note that only a few of her galaxies are low surface 
brightness objects and not all of them \footnote{The only low surface brightness galaxy 
whose CAS parameters have been computed by Conselice is UGC10669, whose clumpiness is nearly zero ($S=-0.07$), as one might expect
from the image.} have been analyzed by Conselice (2003). 
Albeit the measurement method can play a role in this issue, we expect that this affects the measurements by only $\sim 0.1$, 
given the scatter in the right panel of Fig. \ref{comparisonCONS}. We conclude that this discrepancy is not fully understood and 
should be addressed in detail in future studies using a more complete sample of irregular galaxies.

\subsection{The CAS parameter space in the $\rm H\alpha$ line emission}
The concentration index computed from $\rm{H\alpha}$ images $C_{H\alpha}$ is definitely not a function of Hubble type. 
Only a minor decrease can be found from Sd to Im (see Table \ref{tab:CAShatype}). The scatter within each
distribution is larger than the one found for $C_r$. 
Hern{\'a}ndez-Toledo et al. (2008) found that the correlations of the CAS parameters with Hubble type are scattered 
in any SDSS band, but this scatter is greatly reduced in the $r-$ and $i$ bands\footnote{The $z$ band is
found to be intrinsically noisy since the poor filter transmission provides images with lower S/N.},
suggesting that the structure of galaxies is better revealed in the red- and infrared bands.
\begin{table}
\begin{center}
\caption{Average values and $1\sigma$ rms of the CAS parameters measured on $\rm{H\alpha}$ images 
for galaxies with an inclination lower than $66.5^\circ$ on the sky in bins of Hubble type.}
\begin{tabular}{l c c c }
\hline
\hline
Type & $C_{H\alpha}$ & $A_{H\alpha}$ & $S_{H\alpha}$   \\
\hline
Sa - Sb      &  $ 2.2 \pm 0.9 $  & $0.28 \pm 0.15 $ & $ 0.65 \pm 0.30 $ \\   
Sc - Sd      &  $ 2.2 \pm 0.6 $  & $0.29 \pm 0.17 $ & $ 0.69 \pm 0.43 $ \\  
Sdm - Im     &  $ 2.0 \pm 0.9 $  & $0.21 \pm 0.16 $ & $ 0.45 \pm 0.35 $ \\  
BCD          &  $ 2.3 \pm 0.5 $  & $0.36 \pm 0.19 $ & $ 0.33 \pm 0.25 $ \\  
\hline                                                                                                                                                           
\end{tabular}                                              
\label{tab:CAShatype}
\end{center}
\end{table}

As an example we focus on the Sa galaxies. They are bulge-dominated giant spirals whose 
color is redder than that of dwarf star-forming galaxies (Gavazzi et al. 2010). The red color suggests
that they are dominated by old and evolved stars. They show ongoing star formation activity 
that can sometime be quite relevant in an absolute quantity but their star formation rate per unit mass (SSFR) is very low 
compared to the blue dwarfs (Gavazzi et al. 2013a, hereafter Paper II). For this reason their $\rm{H\alpha}$ concentration index 
spans the whole parameter range without any dependence on the morphological classification.
We give two examples of Sa galaxies whose $C_{H\alpha}$ is very different. AGC7538 has a starburst nucleus 
(confirmed both by NED and by the SDSS-DR7 spectrum) and low star formation on the disk so the concentration index 
in the $\rm{H\alpha}$ reaches $C_{H\alpha} = 5.1$. On the other hand, AGC9436 has a giant bulge without star formation 
(the nuclear activity is that of a retired galaxy, Cid Fernandes et al. 2011; Gavazzi et al. 2011) and the star formation activity takes place 
in a thin ring in the disk giving a $C_{H\alpha} = 0.3$ (see also Figure \ref{imaH} bottom).
\begin{figure*}
\begin{center}
\includegraphics[scale=0.60]{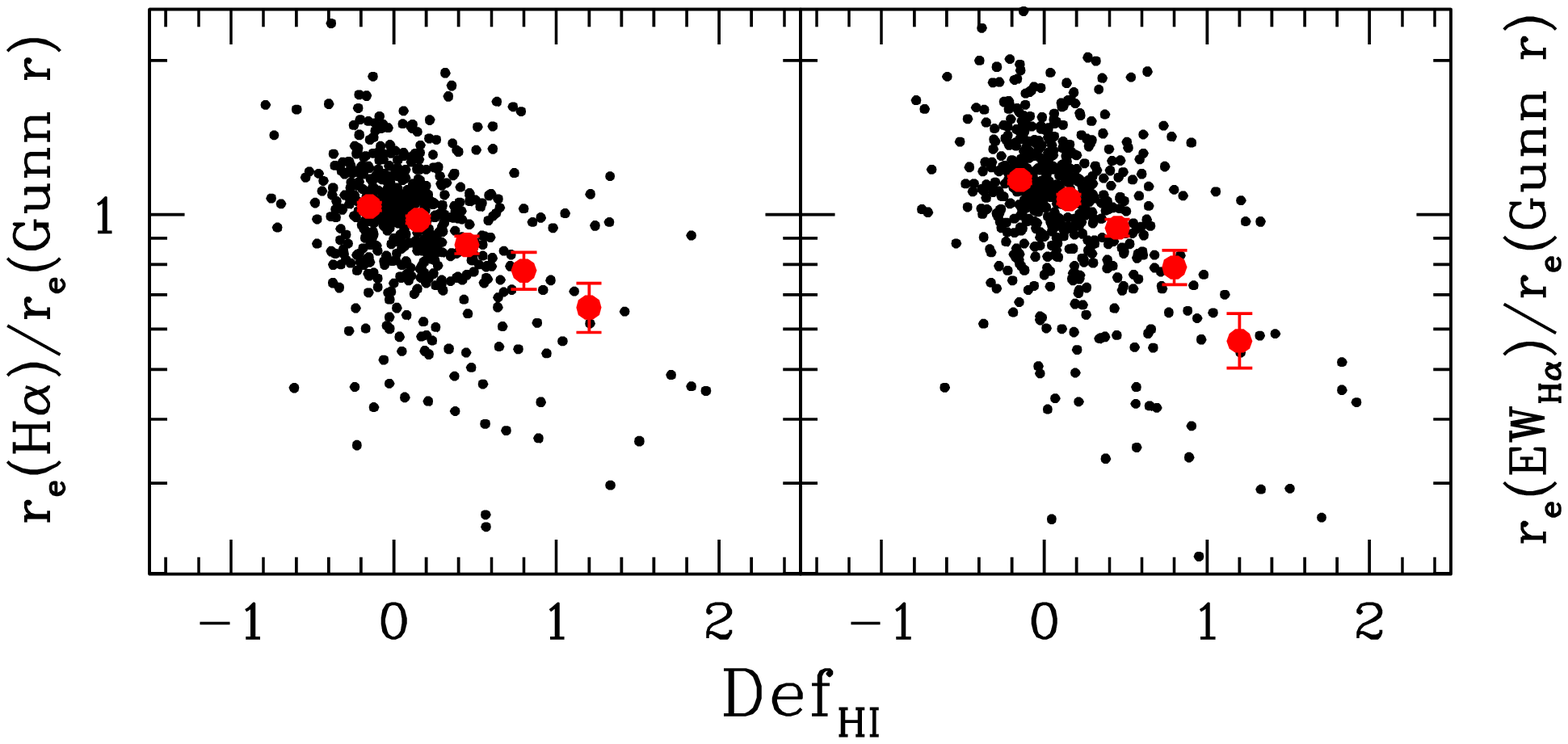}
\caption{Ratio between effective radii (left) and the $EW/r$ parameter (right) as a function of the HI-deficiency.
Big red dots are the average values along the y-axis in different bins of HI deficiency. The $1\sigma$ error bar is 
from bootstrap resampling.}
\label{Ratiodef}
\end{center}
\end{figure*}
The asymmetry in $\rm{H\alpha}$ follows a similarly weak trend as the one computed in the $r$ band (but with larger scatter), 
which supports the theory that the ``flocculent asymmetry'', as it was called by Conselice et al. (2000), is caused by recent 
star formation. For the most part, the contributions to the asymmetry of normal galaxies come from blue 
star-forming regions in the arms of the spirals. 
Finally, the clumpiness in $\rm{H\alpha}$ follows the same trend as a function of Hubble type previously found in the $r$ band,
but the scatter of points within the same bin of type is dramatically greater (see Figure \ref{CAShatype}). 
The scatter is probably due to the stochasticity that affects the star-forming regions especially in the dwarfs 
that are populated by a small number of HII regions (Boselli et al. 2009, Fumagalli et al. 2011, Weisz et al. 2011).

\subsection{Environmental effects on effective radii}

Fig. \ref{Ratiodef} shows the ratio between effective radii (left) and the $EW/r$ parameter (right)  
as a function of the HI-deficiency for all galaxies. The HI-deficiency parameter ($Def_{HI}$) has been defined 
by Haynes \& Giovanelli (1984) as the logarithmic difference between the HI mass observed in one object 
and the expected value in isolated and unperturbed objects of similar type and size. It is commonly used as a proxy for the degree 
of the perturbation that galaxies experience in dense environments. In this work we used an updated calibration 
of the HI-deficiency parameter valid both for giant spirals and dwarf systems (see Paper III).
It is immediately recognizable that environmentally perturbed galaxies have a more concentrated star-formation 
than their old stars ($r_e(\rm{H\alpha}) < r_e(r)$).
This also agrees with the results of Koopmann et al. (2006), who found, on a smaller sample of galaxies in the
Virgo cluster, a similar correlation between HI deficiency and the extension of the $\rm H\alpha$ emission. More 
recently Cortese et al. (2012) have shown a strong correlation between the HI deficiency and the extension of the 
star-forming disk traced by the UV emission. They also performed a comparison between effective and isophotal radii
showing that scatter in the correlation with the HI content can be greatly reduced using isophotal radii from 
radial profile fitting. Unfortunately, as discussed in Sec. \ref{computingCAS}, such technique can not be applied 
safely to $\rm H\alpha$ images. 

It has been shown by Cayatte et al. (1994), Boselli et al. (2006), Fumagalli \& Gavazzi (2008),
Welikala et al. (2008), and Rose et al. (2010) that the HI removal by the ram-pressure stripping (Gunn \& Gott 1972) 
mechanism leads to a truncation of the HI disk. When the HI removal is so severe that the HI disk shrinks inside the optical radius, 
there is also a truncation of the star-forming disk. The efficiency of this process is a function of the environmental
conditions that each galaxy is experiencing (i.e., density and temperature of the intracluster medium and velocity of 
the infalling galaxy). It has been shown both by observations and simulations that the timescale of gas stripping is 
on the order of a few 100 Myr (e.g. Yoshida et al. 2004, 2008, 2012; Boselli et al. 2006, 2008; Kapferer et al. 2009; 
Tonnesen \& Bryan 2009, 2010, 2012; Yagi et al. 2010; Fumagalli et al. 2011).

It is thus recommended that any investigation on the scale at which galaxies produce new stars at $z=0$ should be 
restricted to objects that are currently not undergoing environmental disturbances: i.e., non HI-deficient galaxies.
We set a threshold between deficient and non-deficient galaxies at $def_{HI} = 0.3$, i.e., $1 \sigma$ above the mean
deficiency of the isolated galaxies. The scaling relations presented hereafter are obtained only for non-deficient galaxies.
  
\subsection{Spatial growth of galaxies}

\begin{figure}[ht]
\centering
\includegraphics[width=.94\columnwidth]{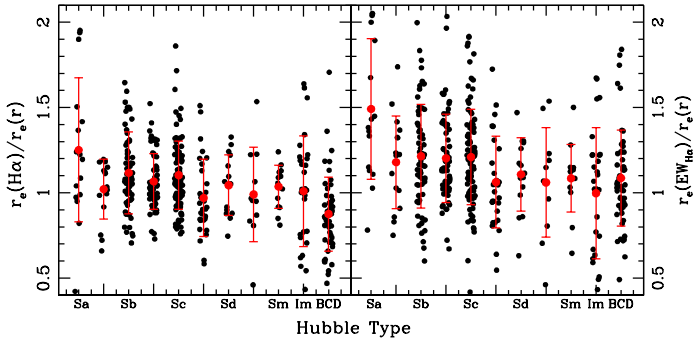}
\caption{Ratio between effective radii (left) and the $EW/r$ parameter (right) as a function of Hubble type. 
The big red dots mark the average ratios and the the $1\sigma$ error bar marks the $1\sigma$ standard deviation.}
\label{Ratiotype}
\end{figure}

Inspecting the ratio between $\rm H\alpha$ and $r$ effective radii as a function of Hubble type, 
we find that the recently born stars ($\rm H\alpha$) generally track the $r$-band stellar continuum, 
but the average values are slightly higher than 1 for giant spirals, consistent with 1 for late spirals and dwarf irregulars, 
and they become lower than 1 for Im and BCDs (see Fig. \ref{Ratiotype}, left panel).

By comparing the $\rm{H\alpha}$ effective radii with those of the stellar continuum, we have two snap-shots: 
a current spatially resolved star formation map, and an age-weighted integral map of the mass built in the past.
These quantities, although they do not directly measure the growth rate of galaxies over time, can be used to infer the growth 
size at the current epoch. Those for which  $r_e({\rm H\alpha}) > r_e(r)$ have a star formation that is more extended than 
their assembled stellar mass, implying that at present they are growing in the outer regions: inside-out growth.
On the other hand, galaxies for which $r_e({\rm H\alpha}) < r_e(r)$ are now building up the mass in the center.
 
\begin{figure*}[tb]
\centering
\includegraphics[width=.90\textwidth]{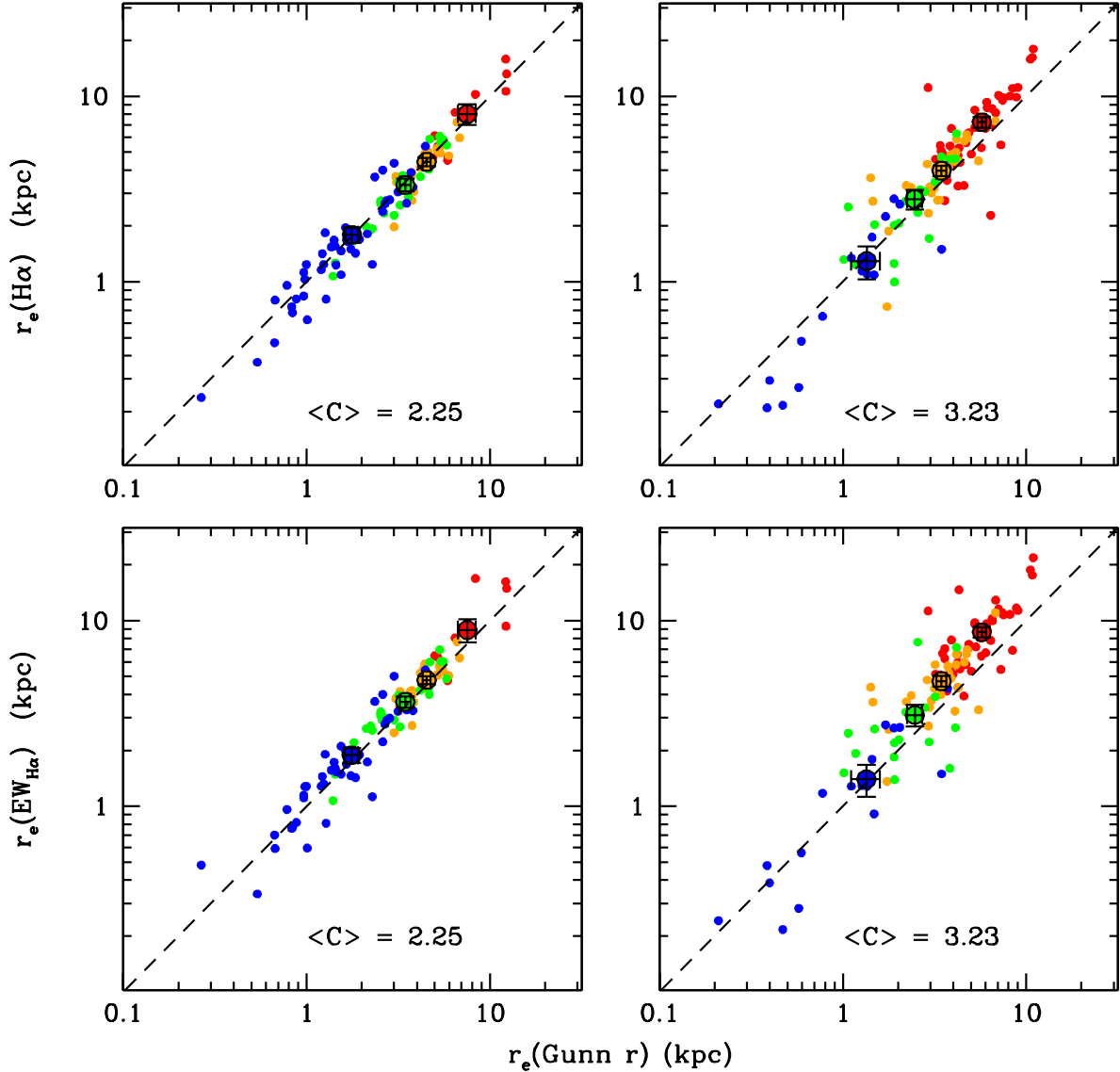}
\caption{Top panels: relation between half-light radii of the $\rm{H\alpha}$ emission and those of the stellar continuum radiation 
in the lowest and highest quartiles of the $r$-band concentration index for unperturbed galaxies. Bottom
panels: the same, but using the half-light radii of the $EW_{\rm{H\alpha}}$ ($EW/r$ parameter). Points are color-coded in 
quartiles of stellar mass: red for ${\rm log}(M_*/M_\odot) \geq 10.3$, orange for  $ 9.8 \leq {\rm log}(M_*/M_\odot) <  10.3$, 
green for $ 9.3 \leq { \rm log}(M_*/M_\odot)<  9.8$, and blue for ${\rm log}(M_*/M_\odot) < 9.3 $. 
The black-contoured big points are the average ratios for each bin of stellar mass.}
\label{RoffHaEWC}
\end{figure*}

Since the morphological classification is intrinsically subjective, we explored possible correlations
between the ratio of the effective radii and the concentration index for unperturbed galaxies.
Fig. \ref{RoffHaEWC} shows  the effective radii of the $\rm{H\alpha}$ emission and those of the stellar continuum 
in the lowest and highest quartiles of the concentration index as measured in the $r$ band. Points are color-coded in quartiles of 
stellar mass and the big points are the average ratios for each bin of stellar mass. 
The average ratios with $1\sigma$ errors from bootstrap resampling are given in Table \ref{RoffnetCt}. 
All averages are consistent with a 1:1 relation except those in the last two bins of concentration and with high 
masses. That is, the most massive and already concentrated galaxies have a star formation that is more extended than 
their stellar mass. This is evidence of an inside-out assembly of disk galaxies at $z\sim0$. 
This result is consistent with those by Nelson et al. (2012) at $z\sim1$  who found an average ratio 
$<r_e({\rm H\alpha}) / r_e(r)>~= 1.3 \pm 0.1$, and with those of Koopmann et al. (2006) at $z\sim0$ who found an
average ratio of $1.14 \pm 0.07$, although this latter result is more uncertain due to small statistics and the use 
of the profile fitting method for the measurements.
These authors also detected a higher ratio  for the largest galaxies, whereas the less massive galaxies of their sample 
$log(M_*/M_\odot)\sim 9.5$ have a mean ratio of one, consistently with our results highlighted in Table 
\ref{RoffnetCt}.

\begin{table*}
\begin{center}
\caption{Average values and $1\sigma$ variations from bootstrap resampling for the ratio
$r_e({\rm H\alpha}) / r_e(r)$ (top) or for the ratio $r_e(EW_{\rm H\alpha}) / r_e(r)$
(bottom) for unperturbed galaxies in quartiles of concentration index 
and stellar mass. No value is given for fewer than 20 objects.}
\begin{tabular}{c c c c c}
\hline
\hline
                               & $ C_r < 2.4 $ &  $ 2.4 < C_r < 2.6 $ & $2.6 <  C_r < 2.9 $ & $ C_r > 2.9 $   \\
\hline
${\rm log}(M_*/M_\odot) < 9.3 $         & $0.99 \pm 0.04$ &  $0.91 \pm 0.04$ & $0.99 \pm 0.07$ & $      -      $  \\
$ 9.3 \leq {\rm log}(M_*/M_\odot)< 9.8$     & $0.94 \pm 0.03$ &  $0.98 \pm 0.03$ & $0.99 \pm 0.03$ & $1.15 \pm 0.09$  \\
$ 9.8 \leq {\rm log}(M_*/M_\odot) < 10.3$   & $0.98 \pm 0.03$ &  $1.01 \pm 0.03$ & $1.10 \pm 0.03$ & $1.19 \pm 0.07$  \\
${\rm log}(M_*/M_\odot) \geq 10.3$          & $      -      $ &  $1.04 \pm 0.03$ & $1.07 \pm 0.03$ & $1.28 \pm 0.07$  \\
\hline  
${\rm log}(M_*/M_\odot) < 9.3 $         & $1.07 \pm 0.04$ &  $1.02 \pm 0.06$ & $1.06 \pm 0.07$ & $      -      $  \\
$ 9.3 \leq {\rm log}(M_*/M_\odot)< 9.8$    & $1.07 \pm 0.03$ &  $1.07 \pm 0.04$ & $1.08 \pm 0.04$ & $1.31 \pm 0.13$  \\
$ 9.8 \leq {\rm log}(M_*/M_\odot) < 10.3$  & $1.07 \pm 0.03$ &  $1.07 \pm 0.04$ & $1.24 \pm 0.04$ & $1.45 \pm 0.09$  \\
${\rm log}(M_*/M_\odot) \geq 10.3$          & $	-       $ &  $1.09 \pm 0.04$ & $1.18 \pm 0.05$ & $1.54 \pm 0.08$  \\
\hline                                                       					                                                                  					            
\end{tabular}                                              
\label{RoffnetCt}
\end{center}
\end{table*}

The newly defined $EW/r$ parameter is less affected by the uncertainties described in Sect. \ref{sizes}.
Computing the effective radius in the EW image makes the central emission of giant spirals 
(which we expect to be the most contaminated by [NII] (Paper I) and dust extinction issues) 
less dominant than the outer disk emission, where the metallicity is lower (Zaritsky et al. 1994; Skillman et al.1996)
and the blue color provides a better continuum removal (Spector et al. 2012). 
Using $EW/r$ increases the trend of the average values for unperturbed galaxies in bins of Hubble type than
when using $r_e(\rm H\alpha)$; (see Fig. \ref{Ratiotype}, right panel). 
The average values are significantly higher than 1 for Sa-Sb spirals, then they decrease toward 1 for later spirals.
We show in the bottom panels of Fig. \ref{RoffHaEWC} the $EW/r$ parameter (i.e., the effective radius computed 
in the EW image as a function of the effective radius in the $r$ band). Less concentrated galaxies have a mean value 
of $EW/r$ of one, regardless of the stellar mass. On the other hand, the most concentrated galaxies have a mean value 
of $EW/r$ that strongly depends on the stellar mass. Dwarf galaxies lie on the 1:1 relation, but with increasing mass 
the points deviate from equality, following a relation slightly steeper than linear.
That is, concentration of the light is the main parameter that influences the growth of galaxies 
at $z\sim 0$, but at fixed concentration the stellar mass plays a second-order role. 
Massive galaxies have built their central bulges at high redshift ($z>1$ Nelson et al. 2012) 
and are now growing in the outer disk. Conversely, the dwarfs are 
still growing on the scale of their previously assembled mass.

\section{Conclusions and summary} \label{conclusect}
We have investigated several structural parameters of $\sim 800$ late-type galaxies in the nearby 
Universe in two optical bands tracing different stellar populations. We used the $\rm{H\alpha}$ 
emission tht traces the youngest stellar population in a galaxy, and the optical $r$ band that traces intermediate-age 
and less massive stars, which constitute the bulk of the mass in galaxies at $z\sim0$. The samples are mainly radio (21cm) 
selected and biased toward gas-rich systems. This selection effect almost completely misses the early types
which were neglected in this work and instead focuses on the most highly star-forming objects in the local Universe,  
covering all late-types, from giants to the low surface brightness irregulars.  The HI reservoir of  
these galaxies is mostly unaffected by environmental mechanisms, and they are the most suitable objects
for studying the effects of secular evolution on the structural parameters.
Nonetheless, we complemented the main HI selection with optically selected galaxies in the densest 
environments to fully understand the effect of the environment on the typical galaxy sizes. 

A reliable statistical analysis of the CAS properties of galaxies requires large, complete, and unbiased samples.
When measuring the CAS parameters in the $r$ band, it appears to be evident that the concentration index depends 
on the stellar mass and on the Hubble type; these variables 
are related because most massive galaxies are bulge dominated, thus most concentrated. Whithin this sequence BCDs 
represent an exception since they are dwarfs but they are as concentrated as intermediate-mass spirals. 
The asymmetry and clumpiness do not exhibit strong correlations either with stellar mass or with Hubble type. 
Although noisy, we detected an increase of these two indices from giant Sa-Sb spirals to Sc-Sd and a subsequent decrement
along the sequence, toward later types. This agrees with Hern{\'a}ndez-Toledo et al. (2008), although they did not consider
dwarf irregulars separately from late spirals. Although even noisier, this pattern is recognizable also in the $A$ and $S$ parameters
measured in the $\rm{H\alpha}$ images. This evidence supports the theory suggested by Conselice et al. (2000) that disregarding
dynamical effects (e.g., tidal interactions or mergers), the asymmetry of a galaxy is dominated by the number and the distribution
of its star-forming regions. The dependence of the CAS parameters on the age of the traced stellar population agrees with
previous findings reported by Taylor-Mager et al. (2007), Hern{\'a}ndez-Toledo et al. (2008), and Fossati et al. (2012).
Our results for the CAS parameters in the $\rm{H\alpha}$ can be compared to those of Taylor-Mager et al. (2007) in the UV since these 
two bands trace similar young stellar populations. 
We found a remarkable agreement for the concentration index, and also for the asymmetry and clumpiness that
are consistent within uncertainties, although it must be taken into account that the statistics of the UV sample used by 
Taylor-Mager et al. (2007) is limited and thus their uncertainties are larger. 
The dependence of the CAS parameters on wavelength is a well-known problem for high-z morphological studies. The optical filters generally used
on the HST probe the UV rest frame at $z\ge1$. There have been attempts to quantify a ``morphological k-correction'' ,
i.e., a conversion factor from a rest-frame UV or optical blue band to the well-known and calibrated optical red bands (Taylor-Mager et al. 2007; 
Conselice et al. 2008). Due to the similar young stellar population traced by $\rm{H\alpha}$ with respect to the UV, our CAS analysis can be used 
as a proxy of the morphology as seen in the ultraviolet bands. Within this framework, the scatter in the $\rm{H\alpha}$ CAS parameters (see Table \ref{tab:CAShatype})
is so severe that finding a proper k-correction as a function of Hubble type is no trivial task. 
Structural properties derived from rest-frame UV bands should be avoided as much as possbile, favoring observations in the IR (rest-frame optical) bands.  

Analyzing the growth in size of galaxies in the Local Universe, we must take the effects of local overdensities into account.
It is well-known that the environment plays a fundamental role in the galaxy evolution. Galaxies perturbed by overdense environments 
have more concentrated recent star formation than their old stars. This is clearly an effect of the truncation of the HI disk, which is the
gas reservoir for the formation of new stars. Although a variety of mechanisms can act together to shape the HI content of a galaxy 
and its star formation in an overdense region (see Boselli \& Gavazzi 2006), combining our results with those in Paper III and taking into 
account the evidence of a gas depletion acting outside-in, we identified ram-pressure stripping (Gunn \& Gott 1972) as the main process involved.
Disentangling the environmental effects from the evolution of isolated and currently unperturbed galaxies, we were also able to investigate
the size at which galaxies are growing at $z\sim0$. Whether the ratio between effective radii in the $\rm{H\alpha}$ and $r$ bands or the $EW/r$ parameter were
used, we found that the concentration index is the main parameter that describes the current growth of galaxies. Less concentrated
objects are now growing at the same scale as their already assembled mass, but for high concentrations the stellar mass plays 
a second-order role. Massive galaxies have built their central mass at high redshift ($z>1$ Nelson et al. 2012) 
and are now growing in the outer disk. This might be expected if gas is accreted onto a galaxy and then cools onto the galaxy 
disk (see e.g. Brooks et al. 2009). However, it is still debated if the size growth of massive galaxies observed from $z\sim2$ down to $z\sim0$ is driven 
by mergers, as maintained by van Dokkum et al. (2010) or simply by the ongoing star formation.  
Conversely, the dwarfs, regardless of their concentration, are now building up their mass on the scale of their old stars.
Recently, P{\'e}rez et al. (2013), using spatially resolved IFU data from the CALIFA (Sanchez et al. 2012) survey, detected a similar 
inside-out growth for local massive galaxies and no gradient in the age of stellar populations for low-mass objects, corroborating the findings
of this work.

\begin{acknowledgements}
The authors acknowledge useful discussion with David J. Wilman, Massimo Dotti, and Emanuele Ripamonti.
The authors thank the anonymous referee for a thorough report, which greatly improved this manuscript, 
and Astrid Peter for his helpful and detailed language revision.
The authors would like to acknowledge the work of the entire ALFALFA collaboration team
in observing, flagging, and extracting the catalog of galaxies used in this work. 
G. G. acknowledges financial support from Italian MIUR PRIN contract 200854ECE5.
Support for M. Fumagalli was provided by NASA through Hubble Fellowship grant HF-51305.01-A awarded by the Space 
Telescope Science Institute, which is operated by the Association of Universities for Research in 
Astronomy, Inc., for NASA, under contract NAS 5-26555.
R.G. and M.P.H. are supported by US NSF grants AST-0607007 and AST-1107390 and by a Brinson Foundation grant.
This research has made extensive use of the GOLDmine database (Gavazzi et al. 2003)
and of the NASA/IPAC Extragalactic Database (NED), which is operated 
by the Jet Propulsion Laboratory, California Institute of Technology, under contract with the
National Aeronautics and Space Administration. 
Funding for the Sloan Digital Sky Survey (SDSS) and SDSS-II has been provided by the 
Alfred P. Sloan Foundation, the Participating Institutions, the National Science Foundation, 
the U.S. Department of Energy, the National Aeronautics and Space Administration, 
the Japanese Monbukagakusho, and 
the Max Planck Society, and the Higher Education Funding Council for England. 
The SDSS Web site is \emph{http://www.sdss.org/}.
The SDSS is managed by the Astrophysical Research Consortium (ARC) for the Participating Institutions. 
The Participating Institutions are the American Museum of Natural History, Astrophysical Institute Potsdam, 
University of Basel, University of Cambridge, Case Western Reserve University, The University of Chicago, 
Drexel University, Fermilab, the Institute for Advanced Study, the Japan Participation Group, 
The Johns Hopkins University, the Joint Institute for Nuclear Astrophysics, the Kavli Institute for 
Particle Astrophysics and Cosmology, the Korean Scientist Group, the Chinese Academy of Sciences (LAMOST), 
Los Alamos National Laboratory, the Max-Planck-Institute for Astronomy (MPIA), the Max-Planck-Institute 
for Astrophysics (MPA), New Mexico State University, Ohio State University, University of Pittsburgh, 
University of Portsmouth, Princeton University, the United States Naval Observatory, and the University 
of Washington.\\
\end{acknowledgements}

\begin{appendix}
\section{Tables of the structural parameters}
Tables A1 and A2 (available only in electronic form at the CDS) contain the structural parameters presented in this work for the Local
and Coma Supercluster galaxies respectively. The data are listed as follows:
\begin{itemize}
\item Col. 1: AGC designation, from Haynes et al. (2011);
\item Col. 2: VCC (Binggeli et al. 1985) designation. For Coma replaced by the CGCG (Zwicky et al. 1968) designation;
\item Cols. 3-4: RA and DEC coordinates (J2000);
\item Cols. 5-7: $r$-band, $\rm{H\alpha}$ and $\rm{EW_{H\alpha}}$ effective radii (in arcsec);
\item Cols. 8-10: concentration, asymmetry and clumpiness (CAS) parameters from $r$-band images;
\item Cols. 11-13: concentration, asymmetry and clumpiness (CAS) parameters from $\rm{H\alpha}$ images.
\end{itemize} 
The effective radii and the CAS parameters from $\rm{H\alpha}$ images are provided only when has been possible to compute them 
(see Sec. \ref{computingCAS} and \ref{sizes}).
\end{appendix}

\end{document}